\documentclass[12pt]{article}

\usepackage[utf8]{inputenc} 
\usepackage[T1]{fontenc}    
\usepackage{hyperref}       
\usepackage{url}            
\usepackage{booktabs}       
\usepackage{amsfonts}       
\usepackage{nicefrac}       
\usepackage{microtype}      
\usepackage{graphicx,psfrag,epsf}
\usepackage{natbib}
\usepackage{doi}
\usepackage{enumerate}
\usepackage{enumitem}

\usepackage{amsmath}
\usepackage{amssymb}
\usepackage{mathtools}
\usepackage{amsthm}
\usepackage{multirow}
\usepackage{makecell}
\usepackage{subcaption}
\usepackage{cleveref}
\usepackage{siunitx}

\theoremstyle{plain}

\newtheorem{theorem}{Theorem}[section]

\theoremstyle{definition}
\newtheorem{definition}[theorem]{Definition}
\theoremstyle{remark}
\newtheorem{remark}[theorem]{Remark}

\newcommand{\norm}[1]{\left\lVert #1 \right\rVert}

\newcommand{\blind}{0}

\begin{document}

\def\spacingset#1{\renewcommand{\baselinestretch}%
{#1}\small\normalsize} \spacingset{1}


\if0\blind
{ \title{\bf  Conditional Mean and Variance Estimation    with Automated Variance Selection via \textit{k}-NN Algorithm}
  \author{
    Marcos Matabuena \\
    Department of Biostatistics \\
    Harvard T.H. Chan School of Public Health \\
    Boston, MA, USA \\
    \\
    Juan C. Vidal \\
    Centro Singular de Investigación en Tecnoloxías Intelixentes \\
    University of Santiago de Compostela \\
    Santiago de Compostela, Spain \\
    \\ 
    Oscar Hernán Madrid Padilla \\
    Department of Statistics \\
    University of California, Los Angeles \\
    Los Angeles, CA, USA \\
    \\
    Jukka-Pekka Onnela \\
    Department of Biostatistics \\
    Harvard T.H. Chan School of Public Health \\
    Boston, MA, USA
  }
  \maketitle
} \fi

\if1\blind
{
  \bigskip
  \bigskip
  \bigskip
  \begin{center}
    {\LARGE\bf  Conditional Mean and Variance Estimation via \textit{k}-NN Algorithm  with Automated Variance Selection}
\end{center}
  \medskip
} \fi
\bigskip
\begin{abstract}
We introduce a novel \textit{k}-nearest neighbor (\textit{k}-NN) regression method for joint estimation of the conditional mean and variance. The proposed algorithm preserves the computational efficiency and manifold-learning capabilities of classical non-parametric \textit{k}-NN models, while integrating a data-driven variable selection step that improves empirical performance. By accurately estimating both conditional mean and variance regression functions, the method effectively reconstructs the conditional distribution and density functions for multiple families of scale-and-localization generative models. We show that our estimator can achieve fast convergence rates, and we derive practical rules for selecting the smoothing parameter~$k$ that enhance the precision of the algorithm in finite sample regimes. Extensive simulations for low, moderate and large-dimensional covariate spaces, together with a real-world biomedical application, demonstrate that the proposed method can consistently outperform the conventional \textit{k-NN} regression algorithm while being more interpretable in the  model output.
\end{abstract}
\noindent%

\spacingset{1.45}

\section{Introduction}
\label{sec:introduction}

Regression analysis and predictive modeling \citep{gyorfi2002distribution} represent the core challenges in both statistics and machine learning. Given a set of predictors \(X\) in a \(p\)-dimensional Euclidean space \(\mathcal{X} = \mathbb{R}^p\) and a response variable of real value \(Y \in \mathbb{R}\), researchers have focused on estimating the mean regression function

\[
m(x) = \mathbb{E}(Y \mid X = x) \quad \text{ for all } x\in \mathcal{X},
\]
\noindent which is central to modeling many scientific and social problems. However, focusing exclusively on the conditional mean ignores the conditional distribution between \(Y\) and \(X\), leading to an incomplete description of the underlying phenomenon \citep{kneib2023rage, klein2024distributional}.

Non-parametric estimation of the full conditional distribution function (and density function),
\[
F(t,x) = \mathbb{P}(Y \le t \mid X=x),
\]
\noindent for each \(t \in \mathbb{R}\) and \(x \in \mathbb{R}^p\), is challenging \citep{hall1999methods, klein2024distributional}. Classical methods such as the Nadaraya-Watson estimator \citep{devroye1978uniform} and local polynomial regression \citep{fan2018local} are heavily affected by the curse of dimensionality \citep{collomb1981estimation} and require strong smoothness assumptions. As a result, many practitioners provided use for parametric models (e.g., linear quantile regression \citep{koenker2005quantile,10.1093/aje/kwu178}, distributional regression methods such as the general framework in \cite{padilla2025risk}, or generalized additive models for location, scale, and shape (GAMLSS) \citep{10.1111/j.1467-9876.2005.00510.x}). In other contexts, such as survival analysis, semiparametric models (e.g., Cox proportional hazards model \citep{cox1972regression} or accelerated failure time models \citep{barnwal2022survival}) are preferred.

Scale location models (SL) provide a useful compromise between nonparametric and simpler parametric models (e.g., linear homoscedastic quantile regression \citep{mu2007power}). SL models assume
\begin{equation}
	Y = m(X) + \epsilon \cdot  \sigma(X),
\end{equation}
\noindent where \(\epsilon\) is a random error that satisfies \(\mathbb{E}(\epsilon)=0\) and \(X \perp \epsilon\) or \(\epsilon \sim \mathcal{N}(0,1)\). Under these conditions, the full model is identifiable, and the functions \(m(\cdot)\) and \(\sigma(\cdot)\) characterize the conditional distribution (and density) in terms of the first and second conditional moments, respectively.

In this paper, we adopt the approach $k$-nearest neighbors (\textit{k}-NN) approach \citep{fix1989discriminatory, stone1977consistent, biau2015lectures} to estimate the conditional mean

\[
m(\cdot) = \mathbb{E}(Y \mid X=\cdot)
\]
\noindent and the conditional variance;
\[
\sigma^2(\cdot) = \operatorname{Var}(Y \mid X=\cdot).
\]

\noindent Our models are designed for large-scale applications and achieve fast convergence rates, particularly when the true regression functions are found in low-dimensional manifolds \citep{kpotufe2011k,vural2018study}.

We also propose a simple yet effective multiple data split strategy to mitigate post-selection bias \citep{chernozhukov2015valid}. In addition, our variable selection methods \citep{guyon2003introduction,bertsimas2020sparse} are adapted to both mean and variance regression functions, substantially enhancing the scalability of the model.


\begin{figure}[ht!]
\centering
\includegraphics[width=0.80\columnwidth]{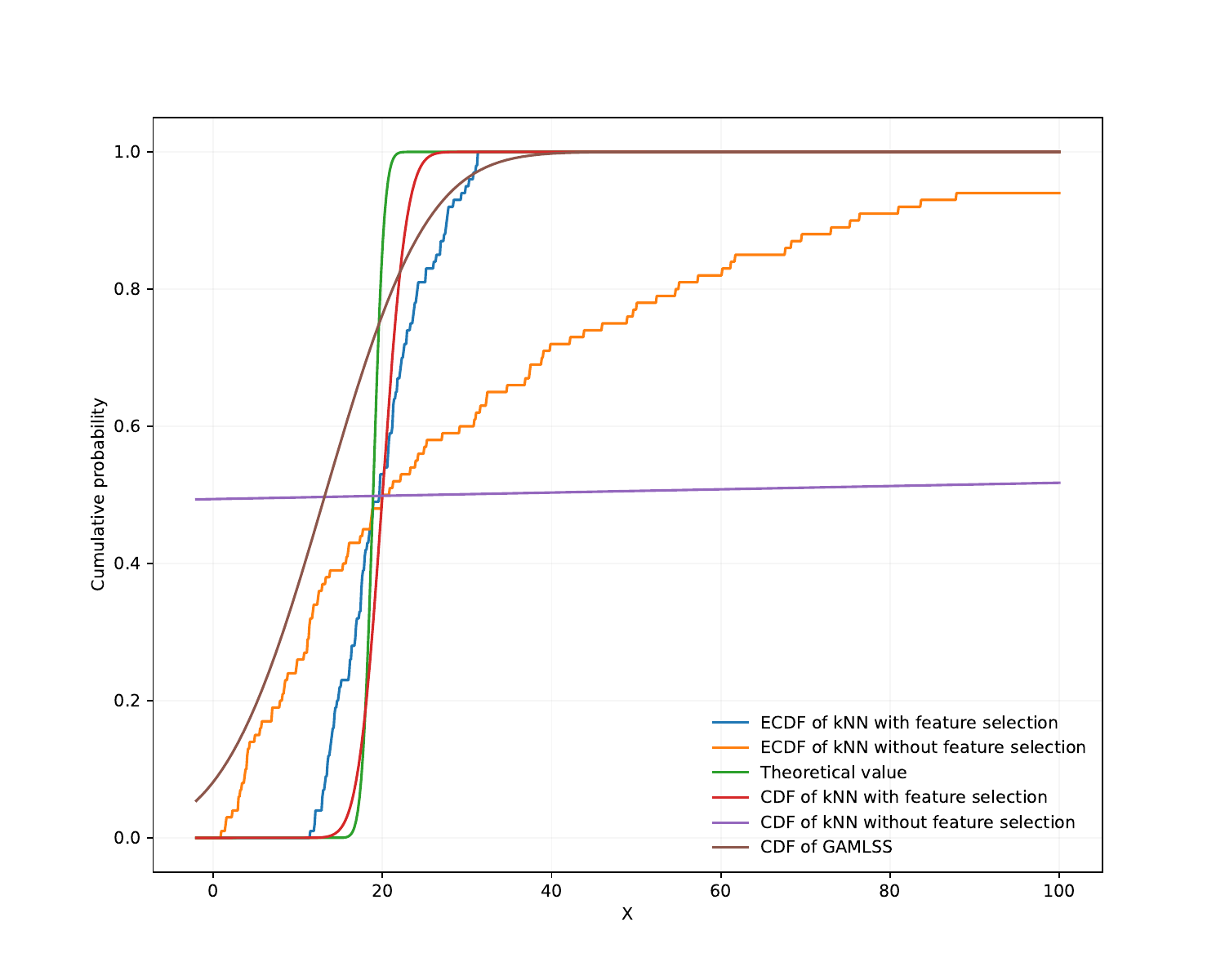}
\vspace{-6mm}
\caption{We estimate the conditional probability $\mathbb{P}(Y \le y \mid X = x)$ for a fixed value $X = x$, using different variants of the $k$‑NN algorithm and GAMLSS. The analysis is based on a specific generative model where $\dim(X) = 10$, but only 4 predictors influence the outcome. The results highlight the importance of appropriate variable selection to accurately approximate the conditional distribution function.}
\label{fig:1}
\vspace{-5mm}
\end{figure}

Variable selection plays a crucial role in our framework for conditional distribution and density estimation and, more broadly, in nonparametric regression, particularly when the true underlying signal lies in a low-dimensional subspace. To illustrate this point, Figure~\ref{fig:1} compares the performance of our method with and without variable selection to approximate the conditional distribution in a one generative example. For additional benchmarking, we also include comparisons with semiparametric GAMLSS models. The results show that omitting variable selection leads to a marked deterioration in model performance.
In addition, the variables that influence the conditional mean function and those that affect the conditional variance function may differ. Identifying the relevant variables for each component improves the performance and interpretability of the model. We also introduce a technically sound criterion for selecting the smoothing parameter \textit{k} (see \citet{zhao2021minimax}) which can have a considerable impact on the estimation of the regression functions.

Our paper is motivated by the need to develop disease risk scores in large biomedical cohorts using $k$NN as a viable and interpretable alternative to traditional parametric and semi-parametric  approaches. In biomedicine, the predominant approach to disease risk prediction is based on parametric models such as linear and logistic regression models. However, recent large data sets from well-characterized clinical populations open the door to incorporating more flexible modeling strategies. Classical non-parametric regressors, such as $k$NN, offer a promising alternative, although they remain underutilized in the biomedical literature. One key advantage of $k$NN is that it does not require strong smoothing assumptions, such as differentiability, which are often imposed by classical models but are rarely verifiable in practice.

\subsection{Summary of Contributions}

Our key contributions are the following.

\begin{enumerate}
    \item \textbf{Efficient and Interpretable Estimation:}
	\begin{itemize}
        \item \textit{Scalability:} The \textit{k}-NN algorithm \citep{andoni2018approximate} provides a scalable framework for handling large datasets, with deriving predictive methods exhibiting quasi-linear complexity.

        \item \textit{Variable Selection and Interpretability:} Our variable selection method improves performance and enhances interpretability by highlighting the impact of key predictors on the mean and variance regression functions.
		
        \item \textit{Theoretical Guarantees:} The \textit{k}-NN variable selection step leads to improved convergence rates in low-dimensional manifold settings \citep{kpotufe2011k}.

        \item \textit{Adaptive Tuning Parameter Selection:} We introduce a data-driven rule to optimally select the parameters \(k\) \citep{azadkia2019optimal}, which improves the performance of the finite sample.
        
        \item \textit{Conditional Distribution and Density Recovery:} Our method accurately recovers the conditional distribution and density in various scale-localization models \citep{akritas2001non}, improving the rates of traditional methods \citep{dombry2023stone}, especially in homoscedastic signal-noise settings \citep{goldfeld1965some}. Based on the previous semiparametric formulation of the model, we propose a prediction interval methodology. 
	\end{itemize}

    \item \textbf{Biomedical Applications:} Our method is a powerful biomedical alternative to large-scale studies with respect to traditional linear or semi-parametric risk models \citep{su2018review}.

\end{enumerate}

\subsection{Outline}

The remainder of the paper is organized as follows. Section~\ref{sec:models} introduces the mathematical models, including regression estimators, variable selection, and prediction interval analysis. Section~\ref{sec:theory} presents the theoretical analysis, including the consistency and convergence rates. Section~\ref{sec:simulation} reports extensive simulation studies in a variety of dimensions and sample sizes. In Section~\ref{sec:real}, we apply our method to biomedical datasets. Finally, Section~\ref{sec:end} concludes with a discussion and future directions.

\section{Background and Related Work}
\label{sec:review}

The \textit{k}-NN algorithm, foundational in non-parametric regression, was theoretically established by \citet{cover1967nearest, stone1977consistent} and practically introduced by \citet{fix1989discriminatory}. Its simplicity and flexibility have led to widespread applications in various predictive tasks \citep{yong2009improved, chen2018explaining, li2021nonparametric}. Unlike smoothers such as the Nadaraya-Watson estimator, \textit{k}-NN does not rely on stringent smoothness assumptions, which is a notable advantage in practice \citep{fan2018local, gyorfi2019nearest}.

\textit{k}-NN performs well with independent and dependent data \citep{biau2010nonparametric} and is effective for continuous and discrete responses \citep{zhang2017efficient}. It has been successfully applied to challenging scenarios such as censored data \citep{chen2019nearest} and counterfactual inference \citep{zhou2017causal}. Moreover, its utility extends to clustering large data sets \citep{shi2018adaptive}, uncertainty quantification \citep{gyorfi2019nearest, lugosi2024uncertainty}, functional data analysis \citep{kara2017data}, and metric space modeling \citep{gyorfi2021universal, cohen2022metric}. Recent theoretical work has advanced adaptive manifold methods \citep{kpotufe2011k, jiang2019non} and minimax estimators \citep{zhao2021minimax}, as well as non-parametric conditional entropy estimation \citep{kozachenko1987sample,10.1214/18-AOS1688}.

Although its traditional focus has been on conditional mean estimation, recent research has increasingly considered conditional variance estimation, especially in low-dimensional settings. This shift was motivated by the development of conditional U statistics \citep{stute1991conditional} and two-step estimation methods \citep{muller1993variance, padilla2022variance}. However, \textit{k}-NN methods have rarely been explored within a semiparametric framework for conditional distribution estimation \citep{dombry2023stone, kosorok2008introduction}, which is a key focus of our work.

Neural networks and deep learning \citep{bartlett2021deep} are often used in large-scale nonparametric regression, yet they face challenges in inference, such as difficulties with bootstrap methods \citep{hardle1988bootstrapping} and convergence to local minima, and are computationally intensive, particularly in high-dimensional settings. In contrast, our \textit{k-NN-based} approach offers a simple and effective computational alternative that demonstrates strong empirical performance in many big data and clinical settings.

\section{Methodology}
\label{sec:models}

Suppose that we observe a random sample

\[
\mathcal{D}_n = \{(X_i, Y_i)\}_{i=1}^{n},
\]

\noindent drawn independently and identically from the joint distribution of \((X, Y) \in \mathcal{X} \times \mathcal{Y}\), where \(\mathcal{X} = \mathbb{R}^p\) and \(\mathcal{Y} = \mathbb{R}\). For simplicity,  we assume \(X = (X^1, \dots, X^p) \sim \mu \),  where $\mu$ is an absolutely continuous probability measure. For analytical purposes, we partition \(\mathcal{D}_n\) into four random subsets \(\mathcal{D}_j\) with the corresponding index set \(\mathcal{S}_j\) of size \(n_j\), for \(j \in \{1,2,3,4\}\).

\subsection{Mathematical Population Framework}

Our objective is to estimate two key regression functions: 
\[
m : \mathbb{R}^{p_1} \rightarrow \mathbb{R} \quad \text{and} \quad \sigma : \mathbb{R}^{p_2} \rightarrow \mathbb{R^{+}},
\]

\noindent which represent the conditional mean and standard deviation regression functions, respectively. To start, we assume only that the functions \( m(\cdot) \) and \( \sigma(\cdot) \) are continuous real-valued functions. However, for rate derivations, we will need to introduce additional smoothness conditions. Here, \(p_1\) and \(p_2\) denote respectively the dimensions of the subspaces that influence the mean and standard deviation regression functions. We assume the following model:
\begin{equation}
	\label{eqn:hom}
	Y = m(X^{\text{mean}}) + \epsilon \cdot \sigma(X^{\text{var}}),
\end{equation}

\noindent where \(\epsilon \sim \mathcal{N}(0,1)\). We define
\[
X^{\text{mean}} = \{X^j : j \in A\} \quad \text{and} \quad X^{\text{var}} = \{X^j : j \in B\},
\]

\noindent with \(A, B \subset [p] = \{1,\dots,p\}\), which contain the indices of variables that influence the mean and standard deviation regression functions. Given $\epsilon\sim \mathcal{N}(0,1)$, the model formulation gives us a full characterization of the conditional distribution function:
\begin{align}
	\label{eqn:dist}
	F(t,x) &= \mathbb{P}(Y \le t \mid X=x) \nonumber \\
	&= \mathbb{P}\Biggl( \epsilon \le \frac{t - m(X^{\text{mean}})}{\sigma(X^{\text{var}})} \,\Biggr|\, X=x \Biggr).
\end{align}

\begin{definition}[Homoscedastic Scale-Localization Model]
The model in \eqref{eqn:hom} is said to be \emph{homoscedastic} if \(\sigma\) is a constant (that is, \(\sigma(\cdot) = c\) for some $c\in \mathbb{R}^{+}$). 
Otherwise, if $\sigma(\cdot)$ is a non-constant function, the model is said to be \emph{heteroscedastic}, which means that the signal-to-noise ratio of the model takes different values across the support of the random variable $X$.

\end{definition}

\paragraph{Prediction Interval Definition.}  
For a new observation \((X_{n+1}, Y_{n+1})\), the oracle prediction interval is the shortest interval satisfying
\[
\mathbb{P}\Bigl(Y_{n+1} \in C^{\alpha}(X_{n+1}) \mid X_{n+1}\Bigr) = 1-\alpha.
\]
 
  \noindent Under the symmetry of the random error $\epsilon$, $C^{\alpha}(\cdot)$ can be written explicitly as
\[
C^{\alpha}(X_{n+1}) = \Bigl[ m(X_{n+1}^{\text{mean}}) - c_\alpha\,\sigma(X_{n+1}^{\text{var}}),\quad m(X_{n+1}^{\text{mean}}) + c_\alpha\,\sigma(X_{n+1}^{\text{var}}) \Bigr],
\]

\noindent where $c_{\alpha}$ is the calibration parameter to ensure conditional nominal coverage $1-\alpha$.

\subsection{Conditional Mean Estimation via \textit{k}-NN Regression}
\label{sec:mean}

We first define the \textit{k}-NN estimator using the whole sample $\mathcal{D}_n$ and then a subsample $\mathcal{D}_j$. Given a positive integer $k_1 := k_{\text{mean}} > 0$ and a norm $\norm{\cdot}$, classify the observations into $\mathcal{D}_n$ by increasing the distance from an arbitrary point $x \in \mathcal{X}$:
 
\[
\norm{x - X_{(1:n)}(x)} \le \cdots \le \norm{x - X_{(n:n)}(x)}.
\]
Define the neighborhood of \(x \in \mathcal{X}\) as
\[
N_{k_1}(x) = \{ i \in [n] : \norm{x - X_i} \le \norm{x - X_{(k_1:n)}(x)} \}.
\]
The \textit{k}-NN estimator for the mean is then
\[
\widehat{m}_{k_1,n}(x) = \frac{1}{k_1} \sum_{i \in N_{k_1}(x)} Y_i.
\]
\noindent For a subsample \(\mathcal{D}_j\), with index set \(\mathcal{S}_j\) (of size \(n_j\)), we define
\[
N_{k_1}^{\mathcal{D}_j}(x) = \{ i \in \mathcal{S}_j : \norm{x - X_i} \le \norm{x - X_{(k_1:n_j)}(x)} \},
\]
 \noindent and the corresponding estimator on the subsample becomes
 
\[
\widehat{m}_{k_1,j}(x) = \frac{1}{k_1} \sum_{i \in N_{k_1}^{\mathcal{D}_j}(x)} Y_i.
\]

\subsection{Conditional Variance Estimation via Residuals}

To estimate the conditional variance, we use a two-step approach. For \(j \geq 2\), after estimating the mean from a previous data split \(\mathcal{D}_{j-1}\), define the residuals for \(i \in \mathcal{S}_j\) by

\begin{equation*}
    \widehat{\epsilon}_i = Y_i - \widehat{m}_{k_1,j-1}(X_i).
\end{equation*}

\noindent The \textit{k}-NN estimator for the variance is
given by

\[
\widehat{\sigma}^2_{k_2,j}(x) = \frac{1}{k_2} \sum_{i \in N_{k_2}^{\mathcal{D}_j}(x)} \widehat{\epsilon}_i^2,
\]
\noindent where \(k_2:= k_{\text{var}} > 0\) and \(N_{k_2}^{\mathcal{D}_j}(x)=\{ i \in \mathcal{S}_j : \norm{x - X_i} \le \norm{x - X_{(k_{2}:n_j)}(x)} \}\). Note that
\[
\operatorname{Var}(Y\mid X=x) = \mathbb{E}[(Y-m(x))^2\mid X=x].
\]

If $\widehat{m}_{k_1,j-1}$ is consistent, then under standard \textit{k}-NN theoretical conditions \citep{stone1977consistent, gyorfi2002distribution, ferrario2018partitioning}, $\widehat{\sigma}^2_{k_2,j}$ is also a consistent estimator.


\subsection{General Variable Selection Strategy for \textit{k}-NN}
\label{sec:var}

Our variable selection procedure is inspired by the explainable ML framework of \citet{verdinelli2021decorrelated}.  
Let \(m(\cdot)\) be the estimator of the mean function, and let  
\(m_{-j}(\cdot)\) denote the same estimator computed after \emph{excluding} the \(j\)-th predictor.

Under the null hypothesis
\[
  H_{0} : \text{``the $j$‑th predictor is not relevant'',}
\]
we consider the loss–difference statistic
\[
  W_{j}(X,Y)
  \;=\;
  \|Y - m_{-j}(X)\|^{2}
  \;-\;
  \|Y - m(X)\|^{2},
\]
and its population counterpart
\[
  w_{j}
  \;=\;
  \mathbb{E}\!\bigl[\|Y - m_{-j}(X)\|^{2}\bigr]
  \;-\;
  \mathbb{E}\!\bigl[\|Y - m(X)\|^{2}\bigr],
  \qquad
  w_{j}\ge 0,\;\;
  w_{j}=0 \; \text{a.s. under } H_{0}.
\]

When simultaneously testing \(d\) predictors, we adjust for multiplicity, for example, through the Bonferroni or false discovery rate control \cite{benjamini1995controlling}.
The empirical analogue of \(w_{j}\) is
\[
  \widetilde{w}_{j}
  \;=\;
  \frac{1}{n_{\ell}}
  \sum_{i\in\mathcal{D}_{\ell}}
  \widetilde{W}_{j}(X_{i},Y_{i}),
  \qquad
  \widetilde{W}_{j}(x,y)
  \;=\;
  \|y-\widehat{m}_{-j,k_{1},n_{\ell}}(x)\|^{2}
  \;-\;
  \|y-\widehat{m}_{k_{1},n_{\ell}}(x)\|^{2},
\]
\noindent where \(\widehat{m}_{k_{1},n_{\ell}}\) is the \textit{k}-NN estimator trained on the complementary training split using \(k_{1}\) neighbors. Due to being evaluated in an independent split, under classical regularity conditions, by \textit{k}-NN  Stone consistency results \cite{stone1977consistent}, and central limit theorem,  
\[
  \sqrt{n_{\ell}}\bigl(\widetilde{w}_{j}-w_{j}\bigr)
  \;\xrightarrow{d}\;
  \mathcal{N}(0,\sigma^{2}_{j}),
\]
\noindent so, the resulting  test statistics is asymptotically normal.


\subsection{Data Splitting Strategy and hyper-parameter selection}
\label{sec:split}

\begin{figure}[bt!]
\centering
\includegraphics[width=0.75\columnwidth]{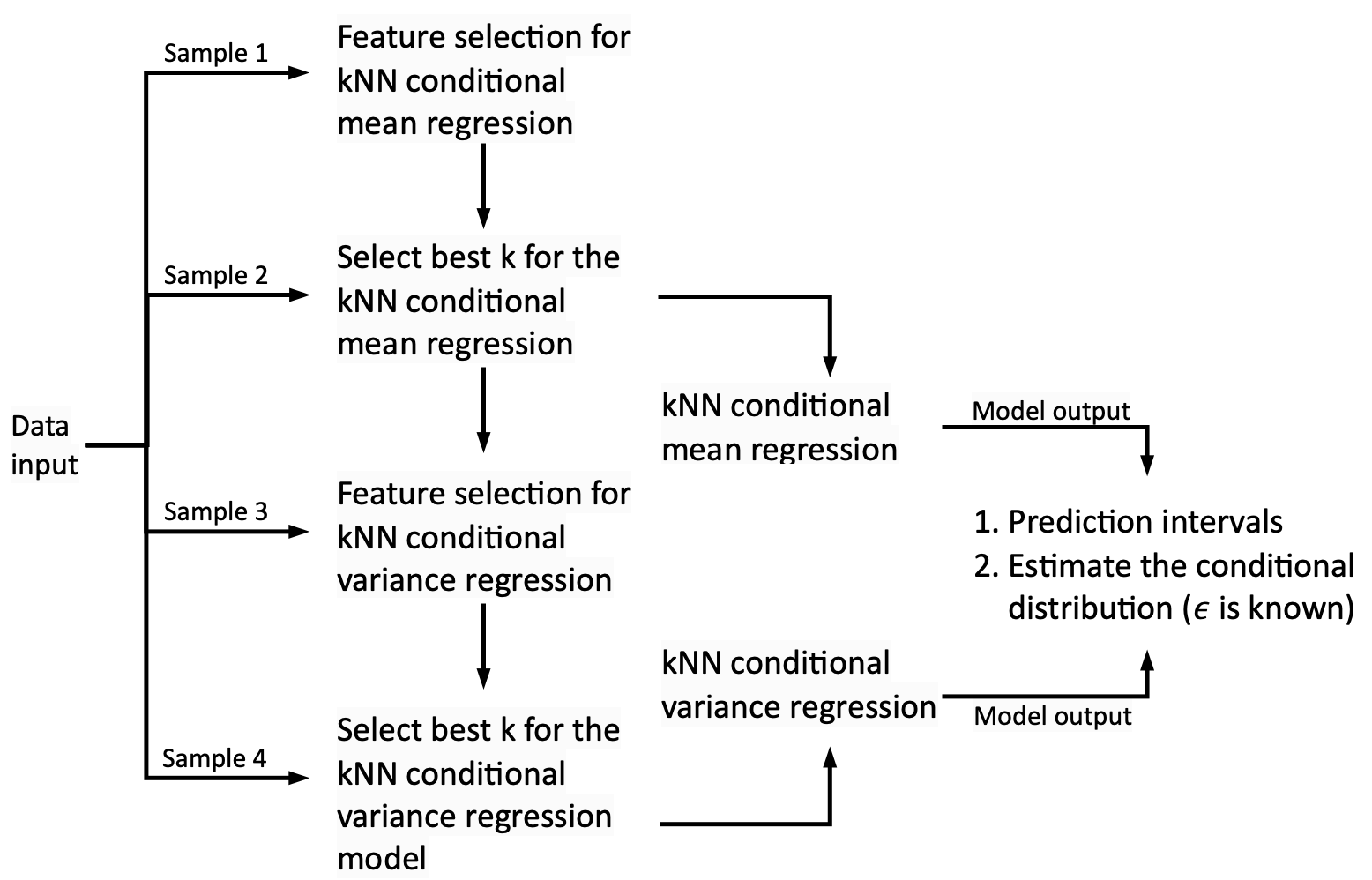}
\caption{Data-splitting strategy and extensions in our \textit{k}-NN semi-parametric framework.}
\label{fig:pipeline}
\end{figure}

To avoid  bias post-selection problems, preserves the gaussian characterization of the test statistics, and obtain computational scalability, we perform variable selection and \(k\) selection on separate data splits at each modeling stage. Figure~\ref{fig:pipeline} illustrates our overall data split strategy, which ensures that our semiparametric algorithm remains robust and computationally efficient even in large datasets. We selected the smoothing parameter $k$ by leave-one-out cross-validation (LOOCV), following the procedure of Azadkia and Chatterjee (2019) \citep{azadkia2019optimal}.

\subsection{Model Extensions: Predictive Interval Algorithm}

Accurate uncertainty quantification is necessary for reliable prediction tasks.
Traditional approaches, such as conformal prediction \citep{shafer2008tutorial, barber2023conformal}, bootstrap methods \citep{zhang2023bootstrap}, and Bayesian inference, have their own computational and model specification limitations.

Recently, \citep{gyorfi2019nearest} introduced a scalable $k$-NN method to construct prediction intervals. To build on this idea but for our  model, we propose two variants: a fully nonparametric version and a semiparametric version.

In the semiparametric case, we first estimate the regression function $\widehat{m}_{k_1,n_{j-2}}$ and the conditional scale function $\widehat{\sigma}_{k_2,n_{j-1}}$ using independent data splits. We then compute standardized residuals
\[
\widehat{\epsilon}_i
= \frac{Y_i - \widehat{m}_{k_1,n_{j-2}}(X_i)}{\widehat{\sigma}_{k_2,n_{j-1}}(X_i)},
\quad i \in \mathcal{S}_{j},
\]
\noindent and form the sample $\{|\widehat{\epsilon}_i|\}_{i\in\mathcal{S}_j}$ to estimate the empirical $(1-\alpha)$-quantile, denoted $\widehat{q}_{1-\alpha}$. The resulting prediction interval is
\[
\widehat{C}^{\alpha}(x)
= \Bigl[
\widehat{m}_{k_1,n_{j-2}}(x) - \widehat{q}_{1-\alpha}\,\widehat{\sigma}_{k_2,n_{j-1}}(x),\;
\widehat{m}_{k_1,n_{j-2}}(x) + \widehat{q}_{1-\alpha}\,\widehat{\sigma}_{k_2,n_{j-1}}(x)
\Bigr].
\]

\noindent Assuming correctly specified Gaussian noise $\epsilon \sim \mathcal{N}(0,1)$, the calibration constant is known in closed form, $q_{1-\alpha}=\Phi^{-1}(1-\alpha/2)$, eliminating the calibration error; the remaining error arises from estimating $\widehat m(
\cdot)$ and $\widehat\sigma(\cdot)$ respectively. This can yield a smaller overall error than fully nonparametric procedures that also estimate the residual distribution. For a new query $x\in\mathbb{R}^p$, the k-NN search typically costs $O(p\log n + k)$ in moderate dimensions with space-partitioning trees, deteriorating to $O(pn)$ in high dimensions; computing the interval from the fitted $\widehat m(\cdot),\widehat\sigma(\cdot)$ is $O(1)$.

\section{Theory}
\label{sec:theory}

In this section, we present our main theoretical results, with detailed proofs deferred to the Supplementary Material. Importantly, for statistical consistency, our assumptions are limited to moment conditions that are always satisfied with clinical biomarkers (given the bounded nature of the variables of interest) and avoid the need for differentiability or smoothness requirements that are difficult to confirm on real data. For the rate deviation and discussion of statistical efficiently respecting a method that involves the conditional distribution function or density based on empirical distribution, we must assume a smoothing condition about the regression functions $m(\cdot)$ and $\sigma(\cdot).$ respectively.

\begin{theorem}[Consistency of the Mean and Variance Regression Functions]
Assume $(\mathbb{E}(Y^{4}) \le L$ for $L>0$. For every $x$ in the support of $\mu$ and every radius $r>0$, we have $\mu\!\bigl(B_r(x)\bigr) > 0$.
 Then, if \(k_1 \to \infty\) with \(k_1/n_1 \to 0\) and \(k_2 \to \infty\) with \(k_2/n_2 \to 0\), the \textit{k}-NN estimators for the mean \(m(\cdot)\) and variance \(\sigma(\cdot)\) are \(L_{2}\)-consistent, that is,
\[
\mathbb{E}\Bigl[ \Bigl| m(X) - \widehat{m}_{k_1,n_1}(X) \Bigr|^{2} \Bigr] \to 0 \quad \text{and} \quad \mathbb{E}\Bigl[ \Bigl| \sigma(X) - \widehat{\sigma}_{k_2,n_2}(X) \Bigr|^{2} \Bigr] \to 0.
\]
\end{theorem}




\begin{remark}
This result extends Stone’s classical consistency theorem for the \(k\)-NN algorithm \citep{stone1977consistent} to the conditional mean and variance estimation based on the conditional mean residuals via data splitting. Unlike traditional nonparametric smoothing methods for these problems, we do not require differentiability assumptions—only \(\mathbb{E}(Y^{4}) < \infty\), which is a mild condition in biomedical applications.

\end{remark}

\begin{theorem}[Rates of the \textit{k}-NN Scale-Localization Gaussian Model]
Let \(\epsilon \sim \mathcal{N}(0,1)\) be as in \eqref{eqn:dist}, and let \(m(\cdot)\) and \(\sigma(\cdot)\) be twice-differentiable regression functions. 
For any fixed \(x \in \mathcal{X}\) and \(y \in \mathcal{Y}\), the error in estimating the conditional distribution function satisfies
\[
\bigl|\widehat{F}(y,x) - F(y,x)\bigr| = O_{\mathbb{P}}\!\bigl(n(m(x))\bigr) + O_{\mathbb{P}}\!\bigl(n(\sigma(x))\bigr),
\]
where \(n(m(x))\) denotes the local convergence rate at the point \(X = x\) for the regression function \(m(\cdot)\), and \(n(\sigma(x))\) the corresponding rate for the standard deviation function \(\sigma(\cdot)\).
\end{theorem}

\begin{remark}
If \(p\) denotes the number of variables influencing the conditional distribution \(F(y,x)\), and these coincide with the variables that affect \(m(x)\) and \(\sigma(x)\), then—up to a constant factor—the rate of the estimator \(k\)-NN for the empirical distribution and our semiparametric approach are equivalent. In contrast, if \(p_1\) or \(p_2\) (the number of relevant variables for \(m(x)\) and \(\sigma(x)\), respectively) is smaller than $|A\cup B|$ (the important variables for the mean and standard deviation regression  simultaneously), our method can automatically achieve faster rates. This justifies the incorporation of variable selection into the \(k\)-NN and semiparametric framework when in the application data lie in a low manifold many times and the variables for the mean and variance regression function can be different and divergent. 
Moreover, for the density estimation of \(f(y,x)\), the semiparametric model always attains a faster rate than the direct nonparametric density estimation, provided the scale-localization (SL) model is specified correctly.
\end{remark}



\begin{theorem}[Universal Consistency of the \textit{k}-NN Variable–Selection Rule]
Fix \(p < n\), and assume that the estimators \(\widehat{m}(\cdot)\) and \(\widehat{\sigma}(\cdot)\) converge uniformly over \(\mathcal{X}\) to the true regression functions \(m(\cdot)\) and \(\sigma(\cdot)\), respectively.  
Let \(k_1, k_2 \to \infty\) with \(k_1 / n_1 \to 0\) and \(k_2 / n_2 \to 0\) as \(n_1, n_2 \to \infty\).  
Then the variable–selection procedure described in Section~\ref{sec:var}, applied to both the mean function \(m(\cdot)\) and the conditional standard deviation function \(\sigma(\cdot)\), defines an \emph{omnibus} test that consistently recovers the true sets of relevant predictors for \(m(\cdot)\) and \(\sigma(\cdot)\); that is,
\[
\mathbb{P}\!\left(\widehat{S}_{m} = S_{m} \;\text{and}\; \widehat{S}_{\sigma} = S_{\sigma}\right) \longrightarrow 1 
\quad \text{as} \quad n_1, n_2 \to \infty ,
\]
\noindent where \(S_{m}, S_{\sigma} \subset [p]\) denote the true sets of indices corresponding to the relevant variables for the mean and variance functions, respectively (denoted earlier as the sets $A$ and $B$ respectively), and \(\widehat{S}_{m}\), \(\widehat{S}_{\sigma}\) denote their estimated counterparts.
\end{theorem}

\begin{remark}
This is not a high-dimensional result: we fix the dimension \(p < n\). 
Nevertheless, it provides an interesting consistency result for many regimes that are commonly satisfied in epidemiological studies and biomedical applications in which larger sample sizes are available. 
It is possible to generalize this result locally, but this requires assuming that the conditional density function is bounded and satisfies additional regularity conditions in order to control the local empirical process of the \(k\)-NN estimator \citep{portier2021nearest}.
\end{remark}


\section{Simulation Study}\label{sec:simulation}

\noindent In this simulation study, we show empirical evidences that the variable‐selection $k$-nearest-neighbors (VS-$k$NN) estimator consistently outperforms both the classical \textit{k}-NN and semi-parametric competitors such as Generalized Additive Models for Location, Scale, and Shape (GAMLSS) \cite{stasinopoulos2017flexible} in estimating the conditional mean, the conditional variance, and in recovering the full conditional distribution. Importantly, VS-$k$NN scales to large datasets: it can be trained on millions of observations in minutes, whereas likelihood-based methods like GAMLSS often become computationally prohibitive even on low-dimensional datasets of approximately 80,000 observations.

\bigskip
\noindent The simulation is designed to answer three key questions:
\begin{enumerate}[label=(\roman*)]
  \item \textbf{Accuracy gain in statistical estimation efficiency from the variable selection step}. How much does VS-$k$NN improve on vanilla \textit{k}-NN error estimation when selecting the most informative features?
  \item \textbf{Comparison with GAMLSS}. To what extent does VS-$k$NN outperform a less flexible semiparametric GAMLSS model in recovering the conditional distribution under nonlinear statistical associations
  \cite{stasinopoulos2017flexible}?
  \item \textbf{Feature selection quality.} To what extent does the proposed variable selection procedure successfully identify the truly active predictorscompared to a standard  Least Absolute Shrinkage and Selection Operator (LASSO)? \cite{tibshirani1996regression}.
\end{enumerate}

In the case of LASSO, the sparsity parameter $\lambda > 0$ is selected via cross-validation, whereas for GAMLSS we use the penalized likelihood approach provided in the GAMLSS R package, which automatically determines the appropriate level of sparsity.

\bigskip

\begin{table}[tb!]
  \centering
  \caption{Simulation scenarios. Each column specifies the conditional mean \( m(X) \), the conditional standard deviation \( \sigma(X) \), and the set of ambient dimensions \( p \) explored.}
  \label{tab:scenarios}
  \begin{tabular}{@{}lcccc@{}}
    \toprule
    \textbf{Regime} & Simulation No. & \(m(X)\) & \(\sigma(X)\) & \(p\) \\
    \midrule
    \multirow[t]{3}{*}{Low} 
        & 1 & \( 5 \cdot (X_2 + 5X_3) \)      & \( 1 \)                        & \( \{3, 10, 20, 25\} \) \\
        & 2 & \( 0 \)                        & \( 5 \cdot X_1 \)              & \( \{3, 10, 20, 25\} \) \\
        & 3 & \( 5 \cdot (X_2 + 5X_3) \)      & \( 5 \cdot X_1 \)              & \( \{3, 10, 20, 25\} \) \\[2pt]
    \multirow[t]{4}{*}{Moderate} 
        & 4 & \( 5 \cdot \sum_{i=1}^{4} X_i \)       & \( 1 \)                        & \( \{5, 10, 20, 50\} \) \\
        & 5 & \( 0 \)                                & \( 5 \cdot \sum_{i=1}^{4} X_i \) & \( \{5, 10, 20, 50\} \) \\
        & 6 & \( 5 \cdot \sum_{i=1}^{3} X_i \)       & \( 5 \cdot (X_4 + X_5) \)        & \( \{5, 10, 20, 50\} \) \\
        & 7 & \( 5 \cdot \sum_{i=1}^{4} X_i \)       & \( 5 \cdot \sum_{i=2}^{5} X_i \) & \( \{5, 10, 20, 50\} \) \\[2pt]
    \multirow[t]{5}{*}{High} 
        & 8  & \( 5 \cdot \sum_{i=1}^{8} X_i \)       & \( 1 \)                            & \( \{10, 25, 50, 100\} \) \\
        & 9  & \( 0 \)                               & \( 5 \cdot \sum_{i=1}^{8} X_i \)   & \( \{10, 25, 50, 100\} \) \\
    \bottomrule
  \end{tabular}
\end{table}

\noindent The nine data-generating scenarios, summarized in Table~\ref{tab:scenarios}, vary the conditional mean function $m(X)$, the conditional standard deviation function $\sigma(X)$, and the ambient dimension $p$ across three regimes (Low, Moderate, High). For each scenario and each of $B = 300$ Monte Carlo replicates, we sample $n$ independent pairs $(X_i, Y_i)$ according to
\[
  Y_i = m(X_i) + \sigma(X_i)\,\varepsilon_i, 
  \quad i\in [n], \quad  \varepsilon_i \sim \mathcal{N}(0,1), 
  \quad X_i \sim \mathrm{Unif}[0,1]^p.
\]

\bigskip


\noindent We explore sample sizes $n \in \{5000, 10000, 20000, 50000, 100000\}$ and ambient dimensions $p \in \{3,5,10,20,25,30,40,50,70,80,100\}$.

\bigskip

\noindent Performance is assessed on an independent test set $\mathcal{D}_{\mathrm{test}}$ of size 5000 using the mean squared error (MSE) for both the conditional mean and conditional standard deviation regression estimators. Denoting by $\widehat m^{(b)}(\cdot)$ and $\widehat \sigma^{(b)}(\cdot)$ the VS-$k$NN estimates in replicate $b\in \{1,\dots,B\}$, we compute the following:
\[
  \widehat{\mathrm{MSE}}_m 
    = \frac{1}{B} \sum_{b=1}^B 
      \frac{1}{|\mathcal{D}_{\mathrm{test}}|} 
      \sum_{x \in \mathcal{D}_{\mathrm{test}}} 
      \bigl[\widehat m^{(b)}(x) - m(x)\bigr]^2,
\]
\[
  \widehat{\mathrm{MSE}}_\sigma 
    = \frac{1}{B} \sum_{b=1}^B 
      \frac{1}{|\mathcal{D}_{\mathrm{test}}|} 
      \sum_{x \in \mathcal{D}_{\mathrm{test}}} 
      \bigl[\widehat \sigma^{(b)}(x) - \sigma(x)\bigr]^2,
\]
\noindent where $\sigma^2(x) = \mathrm{Var}(Y \mid X = x)$.

\subsection{Results: Impact of Variable Selection on kNN}\label{subsec:results}

We begin by exploring how VS-$k$NN behaves across different dimensionality regimes, highlighting Simulations 3, 6 and 9 (Tables~\ref{tab:simul1}, \ref{tab:2} and \ref{tab:3}). The complete results for all settings are provided in Tables~\ref{tab:low1}–\ref{tab:lar2}. Here, we focus only on the main findings. 

\paragraph{Low-dimensional regime ($p \leq 25$).}
When the number of predictors is small, variable selection already makes a noticeable difference. In Scenario 3 ($n=50{,}000$, $p=25$), VS-$k$NN reduces the variance MSE by nearly an order of magnitude compared with vanilla \textit{k}-NN (Table~\ref{tab:simul1}). Interestingly, this improvement in variance control comes without sacrificing accuracy in mean prediction, which stays the same or improves slightly.

\paragraph{Moderate-dimensional regime ($p \leq 50$).}
As the dimensionality increases, the benefits of selection become clearer. In Scenario 6 ($n=100{,}000$, $p=50$), VS-$k$NN cuts the MSE for $\sigma(\cdot)$ by a factor of four and halves the mean MSE relative to \textit{k}-NN (Table~\ref{tab:2}). Here, the selection step proves crucial, effectively filtering out irrelevant variables that otherwise add noise and weaken vanilla \textit{k}-NN.

\paragraph{High-dimensional regime ($p \leq 100$).}
In higher dimensions, the contrast becomes stark. In Scenario 9 ($n=100{,}000$, $p=100$), vanilla \textit{k}-NN essentially breaks down—the MSE increases by more than tenfold. In contrast, VS-$k$NN remains stable, delivering results that are about ten times better for both the conditional mean and the variance (Table~\ref{tab:3}). This robustness suggests that VS-$k$NN not only adapts to moderate noise but also scales gracefully to challenging high-dimensional scenarios.

\subsection{Comparison with GAMLSS}\label{subsec:gamlss}


Table~\ref{tab:gamlss} compares three approaches—VS-\textit{k}-NN with feature selection (FS), VS-\textit{k}-NN without FS, and GAMLSS—in reconstructing the conditional distribution defined by $m(X)=\exp(3*X_1 X_2)$ and $\sigma(X)=X_3+X_4$. Across all sample sizes \(n\in\{5\times10^{3},\,10^{4},\,2\times10^{4},\,5\times10^{4}\}\) and dimensionalities \(p\in\{10,25,50,100\}\), VS-$k$NN with FS consistently achieves the lowest mean squared error (MSE) between the empirical CDF (ECDF) and the GAMLSS normal CDF (NCDF).

\begin{itemize}
  \item \textbf{Magnitude of the gain.} At the smallest setting (\(n=5{,}000,\;p=10\)),  
  \(\mathrm{ECDF_S\text{-}MSE}\) decreases from~0.1068 with GAMLSS to~0.0187 with VS-$k$NN+FS (82\,\% reduction). Similarly, \(\mathrm{NCDF_S\text{-}MSE}\) drops from~0.1068 to~0.0090 (92\,\% reduction).  
  With more data (\(n=50{,}000\)), the gap widens dramatically: \(\mathrm{ECDF_S\text{-}MSE}\) shrinks from~0.1092 to~0.00336 (97\,\% reduction), while \(\mathrm{NCDF_S\text{-}MSE}\) falls from~0.1092 to~0.00048 (99.6\,\% reduction). These gains reflect the fact that VS-$k$NN adapts to the local geometry of the data manifold, whereas GAMLSS relies on a fixed parametric structure that struggles to capture nonlinear interactions at scale.

  \item \textbf{Role of feature selection.} Removing FS substantially degrades statistical accuracy: for instance, at \(n=5{,}000,\;p=10\), \(\mathrm{ECDF_S\text{-}MSE}\) increases fourfold (0.0187~\(\rightarrow\)~0.0795). Similar effects are seen for \(\mathrm{NCDF_S\text{-}MSE}\). This illustrates how irrelevant variables can blur neighborhood relationships, reducing the quality of nonparametric estimates. The variable step is therefore crucial.
  \item \textbf{Stability in dimensions;} Gains relative to GAMLSS remain above 70 \,\% even when \(p=100\), showing that VS-k-NNwith FS scales gracefully to higher dimensions. The method avoids the curse of dimensionality by discarding noise variables, while GAMLSS faces increasing difficulty to adjust the models in larger predictor spaces.
\end{itemize}

Taken together, these findings suggest  the proposed automatic knn  with explicit feature selection not only improves finite-sample efficiency, but also provides robustness across dimensional regimes, allowing VS-$k$NN to outperform a flexible semiparametric baseline in recovering the full conditional distribution.

\subsection{Variable–selection accuracy}

The correct selection of variables can substantially influence the performance of the learning algorithms. The purpose of this subsection is to evaluate the effectiveness of VS-$k$NN in this task. Support-recovery results (Supplementary Tables~\ref{tab:scenario1alpha001}–\ref{tab:scenario9alpha001}) show that the paired loss-difference test underlying VS-$k$NN reliably identifies the active covariates driving both the mean and variance components. In large-dimensional settings (for example, scenario 8 with \(p=100\)), the method achieves perfect accuracy for the mean and at least 0.89 accuracy for the variance component once \(n\ge10{,}000\), reaching 1.00 when \(n\ge20{,}000\). In contrast, LASSO systematically fails to detect predictors of variance only in the same setting, underscoring the importance of explicitly modeling heteroskedasticity. In moderate dimensions (\(p=25\)), VS-$k$NN maintains strong performance, achieving high accuracy even with smaller samples (\(n=5{,}000\)) and improving steadily as the sample size increases. 

In low dimensions (\(p\le25\)), the method consistently recovers the exact support without false positives, even at the smallest sample size considered. Taken together, these findings demonstrate that VS-$k$NN is a powerful variable selection tool for conditional distribution modeling: it preserves power in high-dimensional regimes, provides strict Type I error control, and most importantly, detects variance-only predictors that convex alternatives such as LASSO fail to identify. From a nonparametric perspective, this ability to recover both mean and variance drivers makes VS-$k$NN particularly valuable for uncovering the full structure of conditional distributions in practice.



\begin{table*}[!t]
	\centering
	\begin{tabular}{c|c|cc|cc|cc|cc}
		\multicolumn{2}{c}{$\;$} & \multicolumn{2}{c}{$p=3$} & \multicolumn{2}{c}{$p=10$} & \multicolumn{2}{c}{$p=20$} & \multicolumn{2}{c}{$p=25$}  \\ 
		& $N$ & $\overline{x}$ & $\hat{v}$ & $\overline{x}$ & $\hat{v}$ & $\overline{x}$ & $\hat{v}$ & $\overline{x}$ & $\hat{v}$ \\ 
		\hline
		\parbox[t]{2mm}{\multirow{5}{*}{\rotatebox[origin=c]{90}{FS}}} & 
        5000 & 0.0782 & 0.5546 & 0.0770 & 0.7934 & 0.0741 & 0.9381 & 0.0743 & 0.8808 \\
        & 10000 & 0.0473 & 0.6874 & 0.0482 & 0.6662 & 0.0448 & 0.4978 & 0.0476 & 0.7247 \\
        & 20000 & 0.0313 & 0.4962 & 0.0316 & 0.4151 & 0.0312 & 0.5146 & 0.0310 & 0.5302 \\
        & 50000 & 0.0179 & 0.0583 & 0.0184 & 0.0526 & 0.0167 & 0.0653 & 0.0171 & 0.0519 \\
        & 100000 & 0.0109 & 0.0191 & 0.0114 & 0.0193 & 0.0110 & 0.0180 & 0.0113 & 0.0211 \\
		\hline
		\parbox[t]{2mm}{\multirow{5}{*}{\rotatebox[origin=c]{90}{No FS}}} & 
        5000 & 0.1437 & 1.0584 & 0.7315 & 1.9104 & 1.3806 & 3.3652 & 1.6027 & 4.0641 \\
        & 10000 & 0.0932 & 2.2468 & 0.5888 & 1.6454 & 1.1959 & 3.0601 & 1.4157 & 3.3767 \\
        & 20000 & 0.0643 & 0.8605 & 0.4889 & 1.4443 & 1.0455 & 2.4025 & 1.2613 & 3.0328 \\
        & 50000 & 0.0398 & 0.3016 & 0.3688 & 0.9132 & 0.8894 & 1.7339 & 1.0935 & 2.1656 \\
        & 100000 & 0.0261 & 0.1461 & 0.2963 & 0.6846 & 0.7799 & 1.3983 & 0.9801 & 1.7551 \\
		\hline
	\end{tabular}
	\caption{Performance evaluations in Scenario 3 (low-dimensional regime) are reported for the estimators $\overline{x}$ and $\hat{v}$ under the different VS-$k$NN settings with and without feature selection (denoted FS and No FS, respectively). Results are presented in terms of mean squared error, averaged over 300 Monte Carlo simulations.}
	\label{tab:simul1}
\end{table*}

\begin{table*}[!t]
	\centering
	\begin{tabular}{c|c|cc|cc|cc|cc}
		\multicolumn{2}{c}{$\;$} & \multicolumn{2}{c}{$p=5$} & \multicolumn{2}{c}{$p=10$} & \multicolumn{2}{c}{$p=20$} & \multicolumn{2}{c}{$p=50$}  \\ 
		& $N$ & $\overline{x}$ & $\hat{v}$ & $\overline{x}$ & $\hat{v}$ & $\overline{x}$ & $\hat{v}$ & $\overline{x}$ & $\hat{v}$ \\ 
		\hline
		\parbox[t]{2mm}{\multirow{5}{*}{\rotatebox[origin=c]{90}{FS}}} & 5000 & 0.0184 & 4.7813 & 0.0159 & 5.2681 & 0.0168 & 5.6835 & 0.0168 & 6.0965 \\
		& 10000 & 0.0086 & 4.6833 & 0.0107 & 4.7571 & 0.0091 & 5.9863 & 0.0090 & 5.5643 \\
		& 20000 & 0.0071 & 3.9487 & 0.0063 & 4.3996 & 0.0055 & 4.5909 & 0.0060 & 5.0149 \\
		& 50000 & 0.0294 & 2.2449 & 0.0092 & 2.0947 & 0.0094 & 2.3559 & 0.0091 & 2.4721 \\
		& 100000 & 0.0090 & 0.6105 & 0.0116 & 0.5900 & 0.0064 & 0.6193 & 0.0115 & 0.8261 \\
		
		\hline
		\parbox[t]{2mm}{\multirow{5}{*}{\rotatebox[origin=c]{90}{No FS}}} & 
		5000 & 0.0173 & 4.6294 & 0.0159 & 5.1821 & 0.0167 & 5.6338 & 0.0169 & 6.0650 \\
		& 10000 & 0.0082 & 4.0516 & 0.0106 & 4.5364 & 0.0085 & 5.9468 & 0.0090 & 5.5558 \\
		& 20000 & 0.0059 & 3.2773 & 0.0063 & 4.0511 & 0.0055 & 4.6751 & 0.0059 & 5.3154 \\
		& 50000 & 0.0045 & 1.6746 & 0.0043 & 2.6123 & 0.0045 & 3.5756 & 0.0044 & 4.5143 \\
		& 100000 & 0.0041 & 1.0421 & 0.0042 & 1.9987 & 0.0039 & 2.9740 & 0.0041 & 4.0399 \\
		
		\hline
	\end{tabular}
	\caption{Performance evaluations in scenario 5 (moderate regime)  of the estimators $\overline{x}$ and $\hat{v}$ for the different VS-$k$NN settings described in the text with and without feature selection. We report the mean squared error, averaging over 300 Monte Carlo simulations.}
	\label{tab:2}
\end{table*}

\begin{table*}[!t]
	\centering
	\begin{tabular}{c|c|cc|cc|cc|cc}
		\multicolumn{2}{c}{$\;$} & \multicolumn{2}{c}{$p=10$} & \multicolumn{2}{c}{$p=25$} & \multicolumn{2}{c}{$p=50$} & \multicolumn{2}{c}{$p=100$}  \\ 
		& $N$ & $\overline{x}$ & $\hat{v}$ & $\overline{x}$ & $\hat{v}$ & $\overline{x}$ & $\hat{v}$ & $\overline{x}$ & $\hat{v}$ \\ 
		\hline
		\parbox[t]{2mm}{\multirow{5}{*}{\rotatebox[origin=c]{90}{FS}}} & 
        5000 & 0.9578 & 6.1629 & 1.6626 & 10.9807 & 2.6269 & 14.1726 & 4.0297 & 25.7819 \\
        & 10000 & 0.5283 & 3.2111 & 0.5316 & 3.4864 & 0.5681 & 3.8378 & 0.7101 & 4.4814 \\
        & 20000 & 0.3722 & 2.2244 & 0.3700 & 2.2088 & 0.3737 & 2.5839 & 0.3734 & 2.6613 \\
        & 50000 & 0.2502 & 0.4853 & 0.2496 & 0.4908 & 0.2510 & 0.5408 & 0.2557 & 0.6130 \\
        & 100000 & 0.1892 & 0.2023 & 0.1886 & 0.2080 & 0.1881 & 0.2998 & 0.1914 & 0.4898 \\
		\hline
		\parbox[t]{2mm}{\multirow{5}{*}{\rotatebox[origin=c]{90}{No FS}}} & 
        5000 & 1.7833 & 6.0482 & 3.9554 & 18.5633 & 5.5883 & 34.3815 & 7.0232 & 51.6747 \\
        & 10000 & 1.4065 & 4.6117 & 3.4756 & 14.5157 & 5.1919 & 28.8882 & 6.6454 & 49.9896 \\
        & 20000 & 1.1528 & 3.7377 & 3.0854 & 11.5636 & 4.7953 & 24.9395 & 6.3194 & 41.1798 \\
        & 50000 & 0.8975 & 2.2264 & 2.6725 & 8.1347 & 4.3775 & 19.1781 & 5.9051 & 34.9763 \\
        & 100000 & 0.7144 & 1.6196 & 2.4018 & 6.5065 & 4.0907 & 16.4097 & 5.6321 & 30.7575 \\
		\hline
	\end{tabular}
	\caption{Performance evaluations in scenario 9 (large regime) of the estimators $\overline{x}$ and $\hat{v}$ for the different VS-$k$NN settings described in the text with and without feature selection. We report the mean squared error, averaging over 300 Monte Carlo simulations.}
	\label{tab:3}
\end{table*}

\begin{table}[ht!]
\centering
\begin{tabular}{rrrrrrr}
\toprule
N & P & $\text{ECDF}_{S}$ & $\text{ECDF}_{\bar{S}}$ & $\text{NCDF}_{S}$ & $\text{NCDF}_{\bar{S}}$ & $\text{GAMLSS}$ \\
\midrule
5000 & 10 & 0.018730 & 0.079492 & 0.008994 & 0.190202 & 0.106764 \\
5000 & 25 & 0.018860 & 0.109942 & 0.011472 & 0.239390 & 0.108518 \\
5000 & 50 & 0.018810 & 0.121216 & 0.015254 & 0.240710 & 0.106852 \\
5000 & 100 & 0.018764 & 0.131954 & 0.012136 & 0.240746 & 0.107284 \\
\midrule
10000 & 10 & 0.011776 & 0.068992 & 0.003824 & 0.211980 & 0.106791 \\
10000 & 25 & 0.012496 & 0.099161 & 0.003982 & 0.235362 & 0.111581 \\
10000 & 50 & 0.042173 & 0.113265 & 0.092657 & 0.240032 & 0.108201 \\
10000 & 100 & 0.033683 & 0.129172 & 0.085136 & 0.240855 & 0.108830 \\
\midrule
20000 & 10 & 0.007611 & 0.061436 & 0.002005 & 0.182992 & 0.109465 \\
20000 & 25 & 0.007240 & 0.093247 & 0.001828 & 0.230995 & 0.107175 \\
20000 & 50 & 0.007194 & 0.110255 & 0.002135 & 0.237778 & 0.109285 \\
20000 & 100 & 0.007269 & 0.124131 & 0.001946 & 0.240944 & 0.108272 \\
\midrule
50000 & 10 & 0.003360 & 0.051741 & 0.000478 & 0.154573 & 0.109154 \\
50000 & 25 & 0.003456 & 0.083517 & 0.000507 & 0.230324 & 0.108268 \\
50000 & 50 & 0.003437 & 0.103897 & 0.000605 & 0.237425 & 0.108821 \\
\bottomrule
\end{tabular}

\vspace{2mm}
\caption{Comparison of the MSE between the Empirial and Normal Cumulative Distribution Functions, with ($\text{ECDF}_S$ and $\text{NCDF}_S$) and without the feature selection  ($\text{ECDF}_{\bar{S}}$ and $\text{NCDF}_{\bar{S}}$), obtained from VS-$k$NN approach and the Generalized Additive Models for Location, Scale, and Shape (GAMLSS) approach, for $m(X)=\mathrm{e}^{(3 \cdot X_1 \cdot X_2)}$ and $\sigma(X)=X_3+X_4$. }
\label{tab:gamlss}
\end{table}
\newpage

\section{Clinical Case Study (Fasting Plasma Glucose Prediction in India)}
\label{sec:real}

\begin{figure}[bt!]
\centering
\includegraphics[width=1.0\columnwidth]{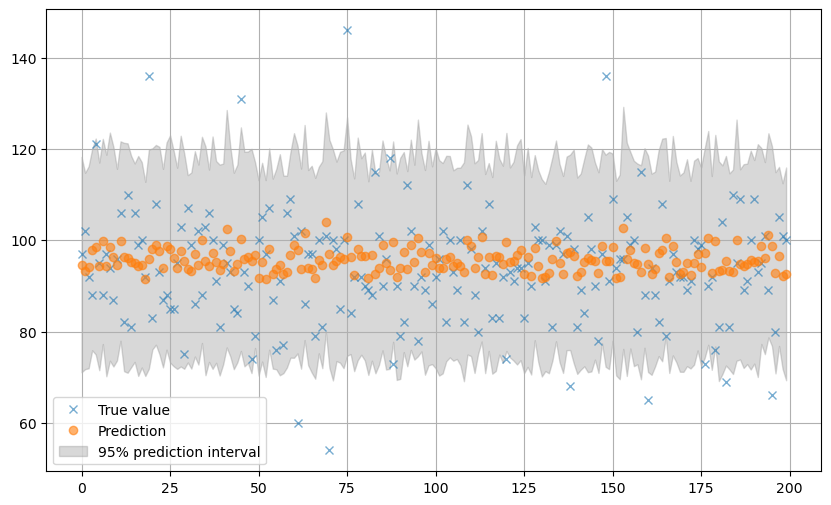}
\caption{Comparison between true values and model predictions over a random sample of 200 participants. Blue crosses denote the true values, while orange circles represent the model’s predictions. The shaded gray area indicates the 95\% prediction interval, constructed from the estimated lower and upper bounds for each prediction. This visualization enables an assessment of both the accuracy and the calibration of the model’s predictive uncertainty.}
\label{fig:bandas}
\end{figure}
In order to illustrate the applicability of the proposed method in real world clinical data, 
we analyze data from the annual Health Survey (AHS) of India (2010--2013), which includes a random sample of 620 \,012 non-diabetic individuals and nine predictor variables. Our goal is to estimate the conditional mean and variance of fasting plasma glucose (FPG) levels and to quantify uncertainty using prediction intervals for FPG, a key biomarker used to diagnose diabetes and monitor disease progression.

 Prediction of FPG can support early detection of diabetes and disease screening, particularly in low-resource settings where risk scores based on individual characteristics guide targeted public health interventions. This is especially relevant in India, which has one of the largest diabetic populations in the world and exhibits substantial socioeconomic heterogeneity.

Table~\ref{tab:resumen} presents summary statistics, including means, standard deviations, maxima, and minima, for the nine predictors and the outcome variable, FPG. For example, the mean age of the sample is \(40.27 \pm 15.96\) years, the mean body mass index (BMI) is \(20.84 \pm 3.41\), and 54\% of the participants are women. The mean level of the outcome FPG is \(95.66 \pm 12.08\) mg / dL, indicating that the population is predominantly normoglycemic according to the criteria of the American Diabetes Association.

Before presenting the regression results for the estimation of conditional mean and variance, Table~\ref{tab:fs-cab} shows the step of selecting variables stratified by gender. The results suggest that anthropometric variables are relevant for predicting the conditional mean of FPG, while to model conditional variability, all variables, except systolic blood pressure, contribute to explaining the uncertainty. This highlights that the set of variables that characterize the conditional mean may differ from those that capture the conditional variance.

\begin{figure}[ht!]
    \centering
    \begin{subfigure}{0.45\textwidth}
        \centering
        \includegraphics[width=\textwidth]{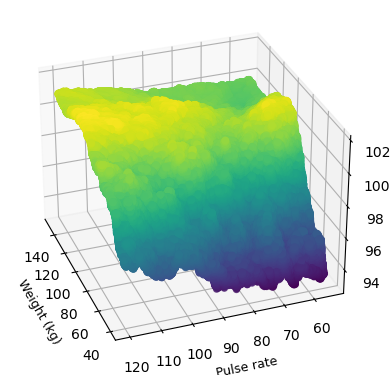}
        \caption{Overall Mean}
        \label{fig:cab-1a}
    \end{subfigure}
    \hfill
    \begin{subfigure}{0.45\textwidth}
        \centering
        \includegraphics[width=\textwidth]{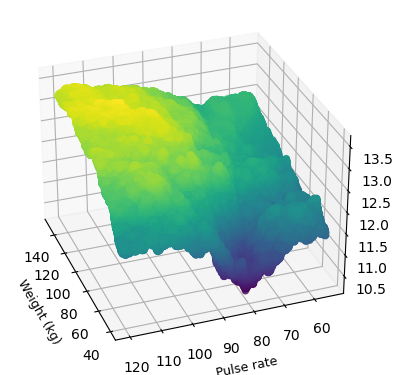}
        \caption{Overall Std. Dev.}
        \label{fig:cab-1v}
    \end{subfigure}
    \vspace{2mm}
    \begin{subfigure}{0.45\textwidth}
        \centering
        \includegraphics[width=\textwidth]{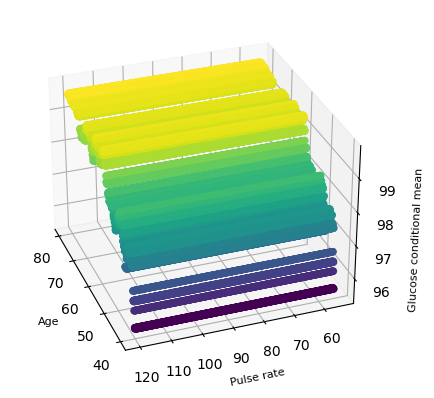}
        \caption{Mean for Men}
        \label{fig:cab-2a}
    \end{subfigure}
    \hfill
    \begin{subfigure}{0.45\textwidth}
        \centering
        \includegraphics[width=\textwidth]{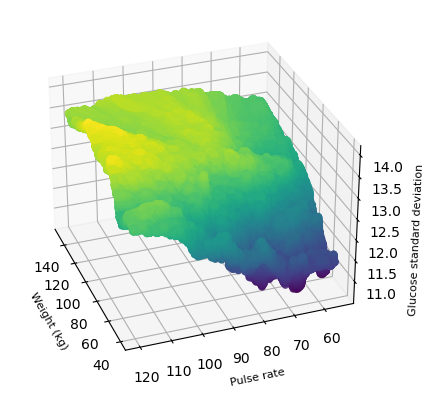}
        \caption{Std. Dev. for Men}
        \label{fig:cab-2v}
    \end{subfigure}
    \caption{Scatter plots showing the predicted FPG for weight and pulse rate among participants in the AHS dataset.}
    \label{fig:cab}
\end{figure}

The scatter plots in Fig.~\ref{fig:cab} that show the results of the conditional mean and variance estimation for different combinations of variance illustrate the influence of gender and individual characteristics in the conditional first and second moments fron the FPG outcome. For example, higher weight and older age in men are associated with an increase in mean FPG, while weight and pulse rate also increase conditional variability. The results obtained emphasize the need for personalized and gender-specific screening strategies for FFG prediction.

Since we did not include any laboratory proxy for glucose values, the model explains only 5\% of the total variance, which underscores the importance of quantifying uncertainty through conditional variance. Figure~\ref{fig:bandas} displays the 95\% prediction intervals for 200 randomly selected individuals. These intervals are relatively wide, reflecting the high uncertainty typical of predictive risk models. In the test set, 93. 9\% of the observations fall within their predicted intervals, indicating a good marginal calibration. Prediction intervals can identify people who fall outside the normal range with a specified confidence level \(\alpha\in(0,1)\) and who may be at greater risk of diabetes.

The algorithm processes more than $6000000$ individuals in less than one minute on a standard desktop, demonstrating its computational efficiency. Future work could explore alternative methods to develop interpretable models that capture additional distributional characteristics as conditional kurtosis beyond means, variances, and prediction intervals. Some conclusions of this analysis are:
i) The sets of variables impacting the conditional mean and variance differ.
ii) Response surfaces can interpret the effects of key predictors even for a predictive disease score for \textit{k}-NN.
iii) Prediction intervals are valuable for identifying uncertainty, identifying outliers, and defining expected normal ranges for clinical outcomes.

\begin{table}
\centering
\caption{Summary of CAB dataset variables. Age (years); Weight (kg); Height (cm); HB, hemoglobin (g/dL); SBP, systolic blood pressure (mmHg); DBP, diastolic blood pressure (mmHg); PR, pulse rate (beats per minute); FBG, fasting blood glucose (mg/dL); BMI, body mass index (kg/m$^2$).}
\label{tab:resumen}
\scalebox{0.85}{
\begin{tabular}{lrrrrrrrrrr}
\toprule
 & AGE & WEIGHT & HEIGHT & HB & SBP & DBP & PR & FBG & BMI \\
\midrule
mean & 40.27 & 50.82 & 156.00 & 9.93 & 122.33 & 77.97 & 80.39 & 95.66 & 20.84 \\
std & 15.96 & 9.70 & 8.56 & 2.32 & 18.52 & 12.83 & 11.53 & 12.08 & 3.41 \\
min & 18.00 & 20.10 & 122.20 & 3.00 & 60.50 & 40.00 & 40.00 & 50.00 & 7.10 \\
max & 99.00 & 236.90 & 188.60 & 18.00 & 249.50 & 160.00 & 140.00 & 154.00 & 104.73 \\
\bottomrule
\end{tabular}}
\end{table}

\begin{table}[ht!]
\centering
\caption{Feature selection in the CAB dataset. Variables include age, weight, height, haemoglobin level, blood pressure measurements, pulse rate, and BMI.}
\label{tab:fs-cab}
\scalebox{0.85}{
\begin{tabular}{@{}lcccccc@{}}
\toprule
& \multicolumn{2}{c}{\textbf{$\;\;\;\;\;$All$\;\;\;\;\;$}}  & \multicolumn{2}{c}{\textbf{$\;\;\;\;$Men$\;\;\;\;$}} & \multicolumn{2}{c}{\textbf{Women}} \\
\cmidrule(lr){2-3} \cmidrule(lr){4-5} \cmidrule(lr){6-7} 
& $\overline{x}$ & $\hat{v}$ & $\overline{x}$ & $\hat{v}$ & $\overline{x}$ & $\hat{v}$ \\ 
\midrule
SEX     &            &            &            &            &            &            \\
AGE     & \checkmark &            & \checkmark & \checkmark & \checkmark & \checkmark \\
WEIGHT  & \checkmark & \checkmark & \checkmark & \checkmark &            &            \\
HEIGHT  &            & \checkmark &            & \checkmark &            & \checkmark \\
HB      & \checkmark & \checkmark & \checkmark & \checkmark &            & \checkmark \\
SBP     &            &            &            &            &            &            \\
DBP     &            & \checkmark &            & \checkmark &            & \checkmark \\
PR      & \checkmark & \checkmark &            & \checkmark &            & \checkmark \\
BMI     &            &  &            &            &            &            \\
\bottomrule
\end{tabular}
}
\end{table}

\section{Discussion and Conclusions}
\label{sec:end}

We have presented a novel, quasi-optimal \textit{k}-NN algorithm incorporating a variable selection step that efficiently approximates conditional distributions in Gaussian-scale and localization models. Our method is theoretically robust and highly scalable, making it well-suited for modern clinical applications where data sets can be extremely large. For example, studies such as those using the UK Biobank \citep{JENKINS2024107} now involve more than 100,000 participants, underscoring the importance of approaches like ours to detect more general statistical associations. Future work will focus on three main directions: i) Extending the algorithm for local variable selection using partition kernel methods \citep{rudi2017falkon}, 
ii) Further investigating the reconstruction of conditional distribution functions under various error models (e.g. beta distributions) and providing a generalization of GAMLSS via \textit{k-NN}-NN, and iii) using our semiparametric approach to define ROC curves in the presence of covariates.

\section{Acknowledgments}
This work received financial support from the Agencia Estatal de Investigación (Spain) (grants PID2023-149549NB-I00, TED2021-130374B-C21, and PID2020-112623GB-I00).

\bibliographystyle{agsm}
\bibliography{biblio}

\bigskip
\begin{center}
{\large\bf SUPPLEMENTAL MATERIALS}
\end{center}

\appendix



\section{Details of \textit{k}-NN}

\subsection{Computational Complexity}

All of the models described in Sections~\ref{sec:mean} to \ref{sec:var} are computationally scalable for large sample sizes. This includes computations for mean, quantile estimations, and inversion of discrete functions. The most computationally intensive part is the \textit{k}-NN algorithm itself, which requires computing pairwise distances \(d_{ij} = \lvert X_i - X_j \rvert\) for \(i,j = 1,\dots,n\).

To improve scalability, we adopt the data-splitting strategy described in Section~\ref{sec:split} and use the Faiss-cpu 1.7.2 library for fast nearest neighbor computations.

\subsection{Software and Data}

\subsubsection{Software and Libraries}
Our \textit{k}-NN model is implemented in Python 3.8.13, using Faiss-cpu 1.7.2 and NumPy 1.22. We use Scikit-learn 1.0.2 for hyperparameter tuning.

\subsubsection{Hardware Specifications}
The calculations were primarily performed on an Intel Xeon Gold 6248 processor (20 cores).

\subsubsection{Reproducibility}
To ensure reproducibility, all random seeds in Python and NumPy were set to 1. Our code, scripts, and preprocessed data sets are available in our public GitHub repository.



\section{Feature Extraction Validation}

Table~\ref{tab:scenario1_5000_alpha001} (and similar tables following) compare the accuracy (ACC) and false positive rate (FPR) of the feature selection performed by the \textit{k}-NN method and Lasso, for both the mean and variance targets, under various settings and with a significance level \(\alpha=0.01\).

\begin{table}[!ht]
\centering
\caption{Comparison of VS-$k$NN and Lasso feature selection accuracy (ACC) and false positive rate (FPR) for the mean and variance in Scenario 1 with \(\alpha = 0.01\).}
\label{tab:scenario1_5000_alpha001}
\scalebox{0.80}{
\begin{tabular}{rrrrrrrrrr}
\toprule
\(N\) & \(P\) & \(\text{ACC}^{\mu}_{k\text{-NN}}\) & \(\text{FPR}^{\mu}_{k\text{-NN}}\) & \(\text{ACC}^{\mu}_{\text{Lasso}}\) & \(\text{FPR}^{\mu}_{\text{Lasso}}\) & \(\text{ACC}^{\sigma^2}_{k\text{-NN}}\) & \(\text{FPR}^{\sigma^2}_{k\text{-NN}}\) & \(\text{ACC}^{\sigma^2}_{\text{Lasso}}\) & \(\text{FPR}^{\sigma^2}_{\text{Lasso}}\) \\
\midrule
5000   & 3  & 1.0000 & 0.0000 & 1.0000 & 0.0000 & 1.0000 & 0.0000 & 0.0000 & 1.0000 \\
5000   & 10 & 1.0000 & 0.0000 & 1.0000 & 0.0000 & 1.0000 & 0.0000 & 0.0000 & 1.0000 \\
5000   & 20 & 1.0000 & 0.0000 & 1.0000 & 0.0000 & 1.0000 & 0.0000 & 0.0000 & 1.0000 \\
5000   & 25 & 1.0000 & 0.0000 & 1.0000 & 0.0000 & 1.0000 & 0.0000 & 0.0000 & 1.0000 \\
\bottomrule
\end{tabular}
}
\end{table}

(Additional tables for Scenarios 2–11 are included similarly.)




\begin{table*}[!t]
	\centering
	\begin{tabular}{c|c|cc|cc|cc|cc}
		\multicolumn{2}{c}{$\;$} & \multicolumn{2}{c}{$p=3$} & \multicolumn{2}{c}{$p=10$} & \multicolumn{2}{c}{$p=20$} & \multicolumn{2}{c}{$p=25$}  \\ 
		& $N$ & $\overline{x}$ & $\hat{v}$ & $\overline{x}$ & $\hat{v}$ & $\overline{x}$ & $\hat{v}$ & $\overline{x}$ & $\hat{v}$ \\ 
		\hline
		\parbox[t]{2mm}{\multirow{5}{*}{\rotatebox[origin=c]{90}{FS}}} & 
        5000 & 0.0434 & 0.0206 & 0.0427 & 0.0117 & 0.0428 & 0.0110 & 0.0421 & 0.0157 \\
        & 10000 & 0.0277 & 0.0242 & 0.0283 & 0.0091 & 0.0280 & 0.0064 & 0.0283 & 0.0105 \\
        & 20000 & 0.0169 & 0.0022 & 0.0171 & 0.0021 & 0.0168 & 0.0024 & 0.0169 & 0.0021 \\
        & 50000 & 0.0096 & 0.0015 & 0.0096 & 0.0015 & 0.0094 & 0.0013 & 0.0096 & 0.0014 \\
        & 100000 & 0.0066 & 0.0012 & 0.0066 & 0.0012 & 0.0066 & 0.0012 & 0.0067 & 0.0012 \\
		
		\hline
		\parbox[t]{2mm}{\multirow{5}{*}{\rotatebox[origin=c]{90}{No FS}}} & 
        5000 & 0.0947 & 0.0222 & 0.6488 & 0.3708 & 1.2960 & 1.4762 & 1.5220 & 2.1521 \\
        & 10000 & 0.0618 & 0.0090 & 0.5156 & 0.2165 & 1.1232 & 1.1237 & 1.3434 & 1.6418 \\
        & 20000 & 0.0405 & 0.0033 & 0.4117 & 0.1316 & 0.9711 & 0.8011 & 1.1863 & 1.2404 \\
        & 50000 & 0.0245 & 0.0019 & 0.3116 & 0.0683 & 0.8116 & 0.4928 & 1.0158 & 0.8055 \\
        & 100000 & 0.0173 & 0.0015 & 0.2568 & 0.0460 & 0.7094 & 0.3606 & 0.9080 & 0.6098 \\	
		\hline
	\end{tabular}
	\caption{Performance evaluations in scenario 1 (low regime)  of the estimators $\overline{x}$ and $\hat{v}$ for the different VS-$k$NN settings described in the text with and without feature selection. We report the mean squared error, averaging over 300 Monte Carlo simulations.}
	\label{tab:low1}
\end{table*}


\begin{table*}[!t]
	\centering
	\begin{tabular}{c|c|cc|cc|cc|cc}
		\multicolumn{2}{c}{$\;$} & \multicolumn{2}{c}{$p=3$} & \multicolumn{2}{c}{$p=10$} & \multicolumn{2}{c}{$p=20$} & \multicolumn{2}{c}{$p=25$}  \\ 
		& $N$ & $\overline{x}$ & $\hat{v}$ & $\overline{x}$ & $\hat{v}$ & $\overline{x}$ & $\hat{v}$ & $\overline{x}$ & $\hat{v}$ \\ 
		\hline
		\parbox[t]{2mm}{\multirow{5}{*}{\rotatebox[origin=c]{90}{FS}}} & 
        5000 & 0.0063 & 0.8026 & 0.0069 & 0.6997 & 0.0061 & 0.9310 & 0.0053 & 1.1172 \\
        & 10000 & 0.0031 & 0.9373 & 0.0025 & 0.6875 & 0.0032 & 0.7145 & 0.0035 & 0.8270 \\
        & 20000 & 0.0022 & 0.5057 & 0.0025 & 0.5198 & 0.0023 & 0.5029 & 0.0019 & 0.5409 \\
        & 50000 & 0.0091 & 0.0595 & 0.0049 & 0.0556 & 0.0032 & 0.0550 & 0.0082 & 0.0604 \\
        & 100000 & 0.0072 & 0.0214 & 0.0064 & 0.0202 & 0.0038 & 0.0204 & 0.0043 & 0.0217 \\
		\hline
		\parbox[t]{2mm}{\multirow{5}{*}{\rotatebox[origin=c]{90}{No FS}}} & 
        5000 & 0.0081 & 1.0415 & 0.0063 & 1.4483 & 0.0066 & 1.6663 & 0.0054 & 1.7433 \\
        & 10000 & 0.0036 & 2.1614 & 0.0037 & 1.3894 & 0.0032 & 1.6022 & 0.0033 & 1.7349 \\
        & 20000 & 0.0022 & 0.8711 & 0.0023 & 1.2565 & 0.0020 & 1.4818 & 0.0020 & 1.5915 \\
        & 50000 & 0.0017 & 0.3060 & 0.0015 & 0.8244 & 0.0015 & 1.1483 & 0.0016 & 1.2290 \\
        & 100000 & 0.0013 & 0.1470 & 0.0014 & 0.6393 & 0.0014 & 0.9694 & 0.0014 & 1.0606 \\
		\hline
	\end{tabular}
	\caption{Performance evaluations in scenario 2 (low regime)  of the estimators $\overline{x}$ and $\hat{v}$ for the different VS-$k$NN settings described in the text with and without feature selection. We report the mean squared error, averaging over 300 Monte Carlo simulations.}
	\label{tab:low2}
\end{table*}


\begin{table*}[!t]
	\centering
	\begin{tabular}{c|c|cc|cc|cc|cc}
		\multicolumn{2}{c}{$\;$} & \multicolumn{2}{c}{$p=3$} & \multicolumn{2}{c}{$p=10$} & \multicolumn{2}{c}{$p=20$} & \multicolumn{2}{c}{$p=25$}  \\ 
		& $N$ & $\overline{x}$ & $\hat{v}$ & $\overline{x}$ & $\hat{v}$ & $\overline{x}$ & $\hat{v}$ & $\overline{x}$ & $\hat{v}$ \\ 
		\hline
		\parbox[t]{2mm}{\multirow{5}{*}{\rotatebox[origin=c]{90}{FS}}} & 
        5000 & 0.0782 & 0.5546 & 0.0770 & 0.7934 & 0.0741 & 0.9381 & 0.0743 & 0.8808 \\
        & 10000 & 0.0473 & 0.6874 & 0.0482 & 0.6662 & 0.0448 & 0.4978 & 0.0476 & 0.7247 \\
        & 20000 & 0.0313 & 0.4962 & 0.0316 & 0.4151 & 0.0312 & 0.5146 & 0.0310 & 0.5302 \\
        & 50000 & 0.0179 & 0.0583 & 0.0184 & 0.0526 & 0.0167 & 0.0653 & 0.0171 & 0.0519 \\
        & 100000 & 0.0109 & 0.0191 & 0.0114 & 0.0193 & 0.0110 & 0.0180 & 0.0113 & 0.0211 \\
		\hline
		\parbox[t]{2mm}{\multirow{5}{*}{\rotatebox[origin=c]{90}{No FS}}} & 
        5000 & 0.1437 & 1.0584 & 0.7315 & 1.9104 & 1.3806 & 3.3652 & 1.6027 & 4.0641 \\
        & 10000 & 0.0932 & 2.2468 & 0.5888 & 1.6454 & 1.1959 & 3.0601 & 1.4157 & 3.3767 \\
        & 20000 & 0.0643 & 0.8605 & 0.4889 & 1.4443 & 1.0455 & 2.4025 & 1.2613 & 3.0328 \\
        & 50000 & 0.0398 & 0.3016 & 0.3688 & 0.9132 & 0.8894 & 1.7339 & 1.0935 & 2.1656 \\
        & 100000 & 0.0261 & 0.1461 & 0.2963 & 0.6846 & 0.7799 & 1.3983 & 0.9801 & 1.7551 \\
		\hline
	\end{tabular}
	\caption{Performance evaluations in scenario 3 (low regime)  of the estimators $\overline{x}$ and $\hat{v}$ for the different VS-$k$NN settings described in the text with and without feature selection. We report the mean squared error,  averaging over 300 Monte Carlo simulations.}
	\label{tab:low3}
\end{table*}

\begin{table*}[!t]
	\centering
	\begin{tabular}{c|c|cc|cc|cc|cc}
		\multicolumn{2}{c}{$\;$} & \multicolumn{2}{c}{$p=5$} & \multicolumn{2}{c}{$p=10$} & \multicolumn{2}{c}{$p=20$} & \multicolumn{2}{c}{$p=50$}  \\ 
		& $N$ & $\overline{x}$ & $\hat{v}$ & $\overline{x}$ & $\hat{v}$ & $\overline{x}$ & $\hat{v}$ & $\overline{x}$ & $\hat{v}$ \\ 
		\hline
		\parbox[t]{2mm}{\multirow{5}{*}{\rotatebox[origin=c]{90}{FS}}} & 
		5000 & 0.1133 & 0.0217 & 0.1141 & 0.0230 & 0.1129 & 0.0347 & 0.1125 & 0.0383 \\
		& 10000 & 0.0813 & 0.0192 & 0.0813 & 0.0110 & 0.0809 & 0.0226 & 0.0818 & 0.0135 \\
		& 20000 & 0.0502 & 0.0042 & 0.0502 & 0.0042 & 0.0499 & 0.0046 & 0.0501 & 0.0044 \\
		& 50000 & 0.0316 & 0.0022 & 0.0316 & 0.0023 & 0.0315 & 0.0023 & 0.0316 & 0.0025 \\
		& 100000 & 0.0205 & 0.0015 & 0.0205 & 0.0015 & 0.0206 & 0.0016 & 0.0205 & 0.0016 \\
		
		\hline
		\parbox[t]{2mm}{\multirow{5}{*}{\rotatebox[origin=c]{90}{No FS}}} & 5000 & 0.3144 & 0.0973 & 0.9331 & 0.7528 & 1.9036 & 3.1338 & 3.2565 & 9.9012 \\
		& 10000 & 0.2196 & 0.0489 & 0.7475 & 0.4774 & 1.6510 & 2.3455 & 2.9914 & 8.2117 \\
		& 20000 & 0.1609 & 0.0229 & 0.5931 & 0.2771 & 1.4356 & 1.7622 & 2.7798 & 7.1134 \\
		& 50000 & 0.1095 & 0.0093 & 0.4424 & 0.1325 & 1.1936 & 1.0660 & 2.5235 & 5.4133 \\
		& 100000 & 0.0776 & 0.0051 & 0.3594 & 0.0865 & 1.0426 & 0.7673 & 2.3573 & 4.5408 \\
		
		\hline
	\end{tabular}
	\caption{Performance evaluations in scenario 4 (moderate regime)  of the estimators $\overline{x}$ and $\hat{v}$ for the different VS-$k$NN settings described in the text with and without feature selection. We report the mean squared error, averaging over 300 Monte Carlo simulations.}
	\label{tab:mod1}
\end{table*}



\begin{table*}[!t]
	\centering
	\begin{tabular}{c|c|cc|cc|cc|cc}
		\multicolumn{2}{c}{$\;$} & \multicolumn{2}{c}{$p=5$} & \multicolumn{2}{c}{$p=10$} & \multicolumn{2}{c}{$p=20$} & \multicolumn{2}{c}{$p=50$}  \\ 
		& $N$ & $\overline{x}$ & $\hat{v}$ & $\overline{x}$ & $\hat{v}$ & $\overline{x}$ & $\hat{v}$ & $\overline{x}$ & $\hat{v}$ \\ 
		\hline
		\parbox[t]{2mm}{\multirow{5}{*}{\rotatebox[origin=c]{90}{FS}}} & 
		5000 & 0.1099 & 2.9097 & 0.1097 & 4.4237 & 0.1122 & 3.9644 & 0.1313 & 3.9069 \\
		& 10000 & 0.0753 & 3.0066 & 0.0730 & 2.9182 & 0.0742 & 4.4914 & 0.0751 & 3.2367 \\
		& 20000 & 0.0451 & 1.9320 & 0.0443 & 1.8292 & 0.0445 & 1.9821 & 0.0452 & 2.1602 \\
		& 50000 & 0.0283 & 0.3856 & 0.0279 & 0.4115 & 0.0280 & 0.4079 & 0.0276 & 0.4491 \\
		& 100000 & 0.0160 & 0.1599 & 0.0160 & 0.1600 & 0.0160 & 0.1666 & 0.0159 & 0.2308 \\
		\hline
		\parbox[t]{2mm}{\multirow{5}{*}{\rotatebox[origin=c]{90}{No FS}}} & 
		5000 & 0.3938 & 2.7810 & 0.8619 & 3.7975 & 1.5102 & 5.7212 & 2.3795 & 14.4896 \\
		& 10000 & 0.2765 & 2.6488 & 0.7041 & 3.1965 & 1.3164 & 5.0554 & 2.1878 & 8.8444 \\
		& 20000 & 0.2127 & 2.1886 & 0.5547 & 2.8820 & 1.1492 & 4.2427 & 2.0193 & 7.2374 \\
		& 50000 & 0.1378 & 1.0577 & 0.4186 & 1.8354 & 0.9508 & 3.0378 & 1.8181 & 5.8541 \\
		& 100000 & 0.1062 & 0.6727 & 0.3475 & 1.3597 & 0.8323 & 2.4507 & 1.6972 & 5.0614 \\
		\hline
	\end{tabular}
	\caption{Performance evaluations in scenario 6 (moderate regime)  of the estimators $\overline{x}$ and $\hat{v}$ for the different VS-$k$NN settings described in the text with and without feature selection. We report the mean squared error, averaging over 300 Monte Carlo simulations.}
	\label{tab:mod3}
\end{table*}



\begin{table*}[!t]
	\centering
	\begin{tabular}{c|c|cc|cc|cc|cc}
		\multicolumn{2}{c}{$\;$} & \multicolumn{2}{c}{$p=5$} & \multicolumn{2}{c}{$p=10$} & \multicolumn{2}{c}{$p=20$} & \multicolumn{2}{c}{$p=50$}  \\ 
		& $N$ & $\overline{x}$ & $\hat{v}$ & $\overline{x}$ & $\hat{v}$ & $\overline{x}$ & $\hat{v}$ & $\overline{x}$ & $\hat{v}$ \\ 
		\hline
		\parbox[t]{2mm}{\multirow{5}{*}{\rotatebox[origin=c]{90}{FS}}} & 
        5000 & 0.4492 & 4.8810 & 0.4748 & 5.7071 & 0.5487 & 6.4132 & 0.8646 & 8.2382 \\
        & 10000 & 0.2070 & 4.3001 & 0.2001 & 4.5830 & 0.2002 & 5.2007 & 0.2037 & 5.6938 \\
        & 20000 & 0.1375 & 4.0639 & 0.1347 & 4.5087 & 0.1361 & 4.7070 & 0.1350 & 5.0836 \\
        & 50000 & 0.0815 & 2.0606 & 0.0806 & 2.1841 & 0.0808 & 2.1381 & 0.0806 & 2.4248 \\
        & 100000 & 0.0581 & 0.6724 & 0.0586 & 0.6185 & 0.0589 & 0.6884 & 0.0583 & 0.8257 \\
		\hline
		\parbox[t]{2mm}{\multirow{5}{*}{\rotatebox[origin=c]{90}{No FS}}} & 
        5000 & 0.5957 & 4.8698 & 1.3052 & 6.4275 & 2.2609 & 11.2373 & 3.5751 & 17.7136 \\
        & 10000 & 0.4200 & 3.9648 & 1.0572 & 5.7865 & 1.9844 & 8.6255 & 3.2899 & 16.4283 \\
        & 20000 & 0.3190 & 3.3921 & 0.8323 & 4.5756 & 1.7203 & 7.3065 & 3.0302 & 14.0230 \\
        & 50000 & 0.2080 & 1.6912 & 0.6273 & 2.9027 & 1.4236 & 5.0885 & 2.7307 & 10.8970 \\
        & 100000 & 0.1584 & 1.0659 & 0.5212 & 2.1141 & 1.2438 & 4.0691 & 2.5403 & 9.3798 \\
		\hline
	\end{tabular}
	\caption{Performance evaluations in scenario 7 (moderate regime) of the estimators $\overline{x}$ and $\hat{v}$ for the different VS-$k$NN settings described in the text with and without feature selection. We report the mean squared error, averaging over 300 Monte Carlo simulations.}
	\label{tab:mod4}
\end{table*}

\begin{table*}[!t]
	\centering
	\begin{tabular}{c|c|cc|cc|cc|cc}
		\multicolumn{2}{c}{$\;$} & \multicolumn{2}{c}{$p=10$} & \multicolumn{2}{c}{$p=25$} & \multicolumn{2}{c}{$p=50$} & \multicolumn{2}{c}{$p=100$}  \\ 
		& $N$ & $\overline{x}$ & $\hat{v}$ & $\overline{x}$ & $\hat{v}$ & $\overline{x}$ & $\hat{v}$ & $\overline{x}$ & $\hat{v}$ \\ 
		\hline
		\parbox[t]{2mm}{\multirow{5}{*}{\rotatebox[origin=c]{90}{FS}}} & 
        5000 & 1.1570 & 1.1374 & 1.8503 & 5.1980 & 3.6125 & 18.6942 & 6.4105 & 49.1656 \\
        & 10000 & 0.8802 & 0.7214 & 0.8812 & 0.6887 & 0.9083 & 0.7995 & 1.0729 & 1.8732 \\
        & 20000 & 0.6726 & 0.3958 & 0.6705 & 0.3862 & 0.6704 & 0.3822 & 0.6718 & 0.3884 \\
        & 50000 & 0.4569 & 0.1633 & 0.4585 & 0.1528 & 0.4577 & 0.1492 & 0.4577 & 0.1496 \\
        & 100000 & 0.3487 & 0.0861 & 0.3492 & 0.0843 & 0.3501 & 0.0852 & 0.3504 & 0.0850 \\
		\hline
		\parbox[t]{2mm}{\multirow{5}{*}{\rotatebox[origin=c]{90}{No FS}}} & 
        5000 & 2.0261 & 3.2376 & 5.1395 & 24.0069 & 7.4416 & 53.6043 & 9.4243 & 84.4730 \\
        & 10000 & 1.6230 & 2.1441 & 4.5736 & 18.7546 & 6.9467 & 44.7917 & 8.9904 & 78.2581 \\
        & 20000 & 1.2911 & 1.3228 & 4.0565 & 14.3289 & 6.4703 & 38.4316 & 8.5534 & 70.1597 \\
        & 50000 & 0.9614 & 0.6103 & 3.4246 & 8.9217 & 5.8054 & 28.6745 & 8.0317 & 57.7030 \\
        & 100000 & 0.7777 & 0.3848 & 3.0665 & 6.8233 & 5.4023 & 24.0036 & 7.7102 & 52.4335 \\
		\hline
	\end{tabular}
	\caption{Performance evaluations in scenario 8 (large regime) of the estimators $\overline{x}$ and $\hat{v}$ for the different VS-$k$NN settings described in the text with and without feature selection. We report the mean squared error, averaging over 300 Monte Carlo simulations.}
	\label{tab:lar1}
\end{table*}


\begin{table*}[!t]
	\centering
	\begin{tabular}{c|c|cc|cc|cc|cc}
		\multicolumn{2}{c}{$\;$} & \multicolumn{2}{c}{$p=10$} & \multicolumn{2}{c}{$p=25$} & \multicolumn{2}{c}{$p=50$} & \multicolumn{2}{c}{$p=100$}  \\ 
		& $N$ & $\overline{x}$ & $\hat{v}$ & $\overline{x}$ & $\hat{v}$ & $\overline{x}$ & $\hat{v}$ & $\overline{x}$ & $\hat{v}$ \\ 
		\hline
		\parbox[t]{2mm}{\multirow{5}{*}{\rotatebox[origin=c]{90}{FS}}} & 
        5000 & 0.9578 & 6.1629 & 1.6626 & 10.9807 & 2.6269 & 14.1726 & 4.0297 & 25.7819 \\
        & 10000 & 0.5283 & 3.2111 & 0.5316 & 3.4864 & 0.5681 & 3.8378 & 0.7101 & 4.4814 \\
        & 20000 & 0.3722 & 2.2244 & 0.3700 & 2.2088 & 0.3737 & 2.5839 & 0.3734 & 2.6613 \\
        & 50000 & 0.2502 & 0.4853 & 0.2496 & 0.4908 & 0.2510 & 0.5408 & 0.2557 & 0.6130 \\
        & 100000 & 0.1892 & 0.2023 & 0.1886 & 0.2080 & 0.1881 & 0.2998 & 0.1914 & 0.4898 \\
		\hline
		\parbox[t]{2mm}{\multirow{5}{*}{\rotatebox[origin=c]{90}{No FS}}} & 
        5000 & 1.7833 & 6.0482 & 3.9554 & 18.5633 & 5.5883 & 34.3815 & 7.0232 & 51.6747 \\
        & 10000 & 1.4065 & 4.6117 & 3.4756 & 14.5157 & 5.1919 & 28.8882 & 6.6454 & 49.9896 \\
        & 20000 & 1.1528 & 3.7377 & 3.0854 & 11.5636 & 4.7953 & 24.9395 & 6.3194 & 41.1798 \\
        & 50000 & 0.8975 & 2.2264 & 2.6725 & 8.1347 & 4.3775 & 19.1781 & 5.9051 & 34.9763 \\
        & 100000 & 0.7144 & 1.6196 & 2.4018 & 6.5065 & 4.0907 & 16.4097 & 5.6321 & 30.7575 \\
		\hline
	\end{tabular}
	\caption{Performance evaluations in scenario 9 (large regime) of the estimators $\overline{x}$ and $\hat{v}$ for the different VS-$k$NN settings described in the text with and without feature selection. We report the mean squared error, averaging over 300 Monte Carlo simulations.}
	\label{tab:lar2}
\end{table*}

To enhance methodological transparency, the following tables \ref{tab:scenario1alpha001} to \ref{tab:scenario9alpha001} —one for each simulation scenario—detail the feature-extraction validation study conducted at the 10\% significance level \(\alpha = 0.1\). For every combination of sample size \(N\) and predictor dimension \(P\), we provide the mean classification accuracy (\(\mathrm{ACC}\)) and the false-positive rate (\(\mathrm{FPR}\)) achieved by the two competing procedures, VS-\(k\)NN and Lasso, when tasked with identifying the true set of informative features governing either the mean (\(\mu\)) or the variance (\(\sigma^{2}\)). This granular breakdown enables a rigorous assessment of each method's robustness to changes in data dimensionality and scale, and furnishes a reproducible benchmark for future comparative investigations.

\begin{table}
\centering
\footnotesize
\begin{tabular}{rrrrrrrrrr}
\toprule
N & P & $\text{ACC}^{\mu}_{k\text{-NN}}$ & $\text{FPR}^{\mu}_{k\text{-NN}}$ & $\text{ACC}^{\mu}_{\text{Lasso}}$ & $\text{FPR}^{\mu}_{\text{Lasso}}$ & $\text{ACC}^{\sigma^2}_{k\text{-NN}}$ & $\text{FPR}^{\sigma^2}_{k\text{-NN}}$ & $\text{ACC}^{\sigma^2}_{\text{Lasso}}$ & $\text{FPR}^{\sigma^2}_{\text{Lasso}}$ \\
\midrule
5000 & 3 & 1.000000 & 0.000000 & 1.000000 & 0.000000 & 1.000000 & 0.000000 & 0.000000 & 1.000000 \\
5000 & 10 & 1.000000 & 0.000000 & 1.000000 & 0.000000 & 1.000000 & 0.000000 & 0.000000 & 1.000000 \\
5000 & 20 & 1.000000 & 0.000000 & 1.000000 & 0.000000 & 1.000000 & 0.000000 & 0.000000 & 1.000000 \\
5000 & 25 & 1.000000 & 0.000000 & 1.000000 & 0.000000 & 1.000000 & 0.000000 & 0.000000 & 1.000000 \\
10000 & 3 & 1.000000 & 0.000000 & 1.000000 & 0.000000 & 1.000000 & 0.000000 & 0.000000 & 1.000000 \\
10000 & 10 & 1.000000 & 0.000000 & 1.000000 & 0.000000 & 1.000000 & 0.000000 & 0.000000 & 1.000000 \\
10000 & 20 & 1.000000 & 0.000000 & 1.000000 & 0.000000 & 1.000000 & 0.000000 & 0.000000 & 1.000000 \\
10000 & 25 & 1.000000 & 0.000000 & 1.000000 & 0.000000 & 1.000000 & 0.000000 & 0.000000 & 1.000000 \\
20000 & 3 & 1.000000 & 0.000000 & 1.000000 & 0.000000 & 1.000000 & 0.000000 & 0.000000 & 1.000000 \\
20000 & 10 & 1.000000 & 0.000000 & 1.000000 & 0.000000 & 1.000000 & 0.000000 & 0.000000 & 1.000000 \\
20000 & 20 & 1.000000 & 0.000000 & 1.000000 & 0.000000 & 1.000000 & 0.000000 & 0.000000 & 1.000000 \\
20000 & 25 & 1.000000 & 0.000000 & 1.000000 & 0.000000 & 1.000000 & 0.000000 & 0.000000 & 1.000000 \\
50000 & 3 & 1.000000 & 0.000000 & 1.000000 & 0.000000 & 1.000000 & 0.000000 & 0.000000 & 1.000000 \\
50000 & 10 & 1.000000 & 0.000000 & 1.000000 & 0.000000 & 1.000000 & 0.000000 & 0.000000 & 1.000000 \\
50000 & 20 & 1.000000 & 0.000000 & 1.000000 & 0.000000 & 1.000000 & 0.000000 & 0.000000 & 1.000000 \\
50000 & 25 & 1.000000 & 0.000000 & 1.000000 & 0.000000 & 1.000000 & 0.000000 & 0.000000 & 1.000000 \\
100000 & 3 & 1.000000 & 0.000000 & 1.000000 & 0.000000 & 1.000000 & 0.000000 & 0.000000 & 1.000000 \\
100000 & 10 & 1.000000 & 0.000000 & 1.000000 & 0.000000 & 1.000000 & 0.000000 & 0.000000 & 1.000000 \\
100000 & 20 & 1.000000 & 0.000000 & 1.000000 & 0.000000 & 1.000000 & 0.000000 & 0.000000 & 1.000000 \\
100000 & 25 & 1.000000 & 0.000000 & 1.000000 & 0.000000 & 1.000000 & 0.000000 & 0.000000 & 1.000000 \\
\bottomrule
\end{tabular}

\vspace{2mm}
\caption{Comparison of VS-$k$NN and Lasso accuracy (ACC) and false positive rate (FPR) for feature selection targeting the Mean and Variance in scenario 1 with $\alpha$ = 0.01.}
\label{tab:scenario1alpha001}
\end{table}

\begin{table}
\centering
\footnotesize
\begin{tabular}{rrrrrrrrrr}
\toprule
N & P & $\text{ACC}^{\mu}_{k\text{-NN}}$ & $\text{FPR}^{\mu}_{k\text{-NN}}$ & $\text{ACC}^{\mu}_{\text{Lasso}}$ & $\text{FPR}^{\mu}_{\text{Lasso}}$ & $\text{ACC}^{\sigma^2}_{k\text{-NN}}$ & $\text{FPR}^{\sigma^2}_{k\text{-NN}}$ & $\text{ACC}^{\sigma^2}_{\text{Lasso}}$ & $\text{FPR}^{\sigma^2}_{\text{Lasso}}$ \\
\midrule
5000 & 3 & 1.000000 & 0.000000 & 1.000000 & 0.000000 & 1.000000 & 0.013400 & 1.000000 & 0.000000 \\
5000 & 10 & 1.000000 & 0.000000 & 1.000000 & 0.000000 & 1.000000 & 0.054000 & 1.000000 & 0.000000 \\
5000 & 20 & 1.000000 & 0.000000 & 1.000000 & 0.000000 & 1.000000 & 0.058000 & 1.000000 & 0.000000 \\
5000 & 25 & 1.000000 & 0.000000 & 1.000000 & 0.000000 & 1.000000 & 0.148600 & 1.000000 & 0.000000 \\
10000 & 3 & 1.000000 & 0.000000 & 1.000000 & 0.000000 & 1.000000 & 0.000000 & 1.000000 & 0.000000 \\
10000 & 10 & 1.000000 & 0.000000 & 1.000000 & 0.000000 & 1.000000 & 0.000000 & 1.000000 & 0.000000 \\
10000 & 20 & 1.000000 & 0.000000 & 1.000000 & 0.000000 & 1.000000 & 0.010000 & 1.000000 & 0.000000 \\
10000 & 25 & 1.000000 & 0.000000 & 1.000000 & 0.000000 & 1.000000 & 0.053400 & 1.000000 & 0.000000 \\
20000 & 3 & 1.000000 & 0.000000 & 1.000000 & 0.000000 & 1.000000 & 0.000000 & 1.000000 & 0.000000 \\
20000 & 10 & 1.000000 & 0.000000 & 1.000000 & 0.000000 & 1.000000 & 0.000000 & 1.000000 & 0.000000 \\
20000 & 20 & 1.000000 & 0.000000 & 1.000000 & 0.000000 & 1.000000 & 0.000000 & 1.000000 & 0.000000 \\
20000 & 25 & 1.000000 & 0.000000 & 1.000000 & 0.000000 & 1.000000 & 0.020000 & 1.000000 & 0.000000 \\
50000 & 3 & 1.000000 & 0.000000 & 1.000000 & 0.000000 & 1.000000 & 0.000000 & 1.000000 & 0.000000 \\
50000 & 10 & 1.000000 & 0.000000 & 1.000000 & 0.000000 & 1.000000 & 0.000000 & 1.000000 & 0.000000 \\
50000 & 20 & 1.000000 & 0.000000 & 1.000000 & 0.000000 & 1.000000 & 0.020000 & 1.000000 & 0.000000 \\
50000 & 25 & 1.000000 & 0.000000 & 1.000000 & 0.000000 & 1.000000 & 0.000000 & 1.000000 & 0.000000 \\
100000 & 3 & 1.000000 & 0.000000 & 1.000000 & 0.000000 & 1.000000 & 0.000000 & 1.000000 & 0.000000 \\
100000 & 10 & 1.000000 & 0.000000 & 1.000000 & 0.000000 & 1.000000 & 0.000000 & 1.000000 & 0.000000 \\
100000 & 20 & 1.000000 & 0.000000 & 1.000000 & 0.000000 & 1.000000 & 0.000000 & 1.000000 & 0.000000 \\
100000 & 25 & 1.000000 & 0.000000 & 1.000000 & 0.000000 & 1.000000 & 0.010000 & 1.000000 & 0.000000 \\
\bottomrule
\end{tabular}

\vspace{2mm}
\caption{Comparison of VS-$k$NN and Lasso accuracy (ACC) and false positive rate (FPR) for feature selection targeting the Mean and Variance in scenario 2 with $\alpha$ = 0.01.}
\label{tab:scenario2alpha001}
\end{table}

\begin{table}
\centering
\footnotesize
\begin{tabular}{rrrrrrrrrr}
\toprule
N & P & $\text{ACC}^{\mu}_{k\text{-NN}}$ & $\text{FPR}^{\mu}_{k\text{-NN}}$ & $\text{ACC}^{\mu}_{\text{Lasso}}$ & $\text{FPR}^{\mu}_{\text{Lasso}}$ & $\text{ACC}^{\sigma^2}_{k\text{-NN}}$ & $\text{FPR}^{\sigma^2}_{k\text{-NN}}$ & $\text{ACC}^{\sigma^2}_{\text{Lasso}}$ & $\text{FPR}^{\sigma^2}_{\text{Lasso}}$ \\
\midrule
5000 & 3 & 1.000000 & 0.000000 & 1.000000 & 0.000000 & 1.000000 & 0.026800 & 0.000000 & 1.000000 \\
5000 & 10 & 1.000000 & 0.000000 & 1.000000 & 0.000000 & 1.000000 & 0.082000 & 0.000000 & 1.000000 \\
5000 & 20 & 1.000000 & 0.000000 & 1.000000 & 0.000000 & 1.000000 & 0.038000 & 0.000000 & 1.000000 \\
5000 & 25 & 1.000000 & 0.000000 & 1.000000 & 0.000000 & 1.000000 & 0.234600 & 0.000000 & 1.000000 \\
10000 & 3 & 1.000000 & 0.000000 & 1.000000 & 0.000000 & 1.000000 & 0.000000 & 0.000000 & 1.000000 \\
10000 & 10 & 1.000000 & 0.000000 & 1.000000 & 0.000000 & 1.000000 & 0.000000 & 0.000000 & 1.000000 \\
10000 & 20 & 1.000000 & 0.000000 & 1.000000 & 0.000000 & 1.000000 & 0.010000 & 0.000000 & 1.000000 \\
10000 & 25 & 1.000000 & 0.000000 & 1.000000 & 0.000000 & 1.000000 & 0.030000 & 0.000000 & 1.000000 \\
20000 & 3 & 1.000000 & 0.000000 & 1.000000 & 0.000000 & 1.000000 & 0.000000 & 0.000000 & 1.000000 \\
20000 & 10 & 1.000000 & 0.000000 & 1.000000 & 0.000000 & 1.000000 & 0.000000 & 0.000000 & 1.000000 \\
20000 & 20 & 1.000000 & 0.000000 & 1.000000 & 0.000000 & 1.000000 & 0.000000 & 0.000000 & 1.000000 \\
20000 & 25 & 1.000000 & 0.000000 & 1.000000 & 0.000000 & 1.000000 & 0.010000 & 0.000000 & 1.000000 \\
50000 & 3 & 1.000000 & 0.000000 & 1.000000 & 0.000000 & 1.000000 & 0.000000 & 0.000000 & 1.000000 \\
50000 & 10 & 1.000000 & 0.000000 & 1.000000 & 0.000000 & 1.000000 & 0.000000 & 0.000000 & 1.000000 \\
50000 & 20 & 1.000000 & 0.000000 & 1.000000 & 0.000000 & 1.000000 & 0.010000 & 0.000000 & 1.000000 \\
50000 & 25 & 1.000000 & 0.000000 & 1.000000 & 0.000000 & 1.000000 & 0.010000 & 0.000000 & 1.000000 \\
100000 & 3 & 1.000000 & 0.000000 & 1.000000 & 0.000000 & 1.000000 & 0.000000 & 0.000000 & 1.000000 \\
100000 & 10 & 1.000000 & 0.000000 & 1.000000 & 0.000000 & 1.000000 & 0.000000 & 0.000000 & 1.000000 \\
100000 & 20 & 1.000000 & 0.000000 & 1.000000 & 0.000000 & 1.000000 & 0.000000 & 0.000000 & 1.000000 \\
100000 & 25 & 1.000000 & 0.000000 & 1.000000 & 0.000000 & 1.000000 & 0.020000 & 0.000000 & 1.000000 \\
\bottomrule
\end{tabular}

\vspace{2mm}
\caption{Comparison of VS-$k$NN and Lasso accuracy (ACC) and false positive rate (FPR) for feature selection targeting the Mean and Variance in scenario 3 with $\alpha$ = 0.01.}
\label{tab:scenario3alpha001}
\end{table}

\begin{table}
\centering
\footnotesize
\begin{tabular}{rrrrrrrrrr}
\toprule
N & P & $\text{ACC}^{\mu}_{k\text{-NN}}$ & $\text{FPR}^{\mu}_{k\text{-NN}}$ & $\text{ACC}^{\mu}_{\text{Lasso}}$ & $\text{FPR}^{\mu}_{\text{Lasso}}$ & $\text{ACC}^{\sigma^2}_{k\text{-NN}}$ & $\text{FPR}^{\sigma^2}_{k\text{-NN}}$ & $\text{ACC}^{\sigma^2}_{\text{Lasso}}$ & $\text{FPR}^{\sigma^2}_{\text{Lasso}}$ \\
\midrule
5000 & 5 & 1.000000 & 0.000000 & 1.000000 & 0.000000 & 1.000000 & 0.000000 & 0.000000 & 1.000000 \\
5000 & 10 & 1.000000 & 0.000000 & 1.000000 & 0.000000 & 1.000000 & 0.000000 & 0.000000 & 1.000000 \\
5000 & 20 & 1.000000 & 0.000000 & 1.000000 & 0.000000 & 1.000000 & 0.000000 & 0.000000 & 1.000000 \\
5000 & 25 & 1.000000 & 0.000000 & 1.000000 & 0.000000 & 1.000000 & 0.000000 & 0.000000 & 1.000000 \\
10000 & 5 & 1.000000 & 0.000000 & 1.000000 & 0.000000 & 1.000000 & 0.000000 & 0.000000 & 1.000000 \\
10000 & 10 & 1.000000 & 0.000000 & 1.000000 & 0.000000 & 1.000000 & 0.000000 & 0.000000 & 1.000000 \\
10000 & 20 & 1.000000 & 0.000000 & 1.000000 & 0.000000 & 1.000000 & 0.000000 & 0.000000 & 1.000000 \\
10000 & 25 & 1.000000 & 0.000000 & 1.000000 & 0.000000 & 1.000000 & 0.000000 & 0.000000 & 1.000000 \\
20000 & 5 & 1.000000 & 0.000000 & 1.000000 & 0.000000 & 1.000000 & 0.000000 & 0.000000 & 1.000000 \\
20000 & 10 & 1.000000 & 0.000000 & 1.000000 & 0.000000 & 1.000000 & 0.000000 & 0.000000 & 1.000000 \\
20000 & 20 & 1.000000 & 0.000000 & 1.000000 & 0.000000 & 1.000000 & 0.000000 & 0.000000 & 1.000000 \\
20000 & 25 & 1.000000 & 0.000000 & 1.000000 & 0.000000 & 1.000000 & 0.000000 & 0.000000 & 1.000000 \\
50000 & 5 & 1.000000 & 0.000000 & 1.000000 & 0.000000 & 1.000000 & 0.000000 & 0.000000 & 1.000000 \\
50000 & 10 & 1.000000 & 0.000000 & 1.000000 & 0.000000 & 1.000000 & 0.000000 & 0.000000 & 1.000000 \\
50000 & 20 & 1.000000 & 0.000000 & 1.000000 & 0.000000 & 1.000000 & 0.000000 & 0.000000 & 1.000000 \\
50000 & 25 & 1.000000 & 0.000000 & 1.000000 & 0.000000 & 1.000000 & 0.000000 & 0.000000 & 1.000000 \\
100000 & 5 & 1.000000 & 0.000000 & 1.000000 & 0.000000 & 1.000000 & 0.000000 & 0.000000 & 1.000000 \\
100000 & 10 & 1.000000 & 0.000000 & 1.000000 & 0.000000 & 1.000000 & 0.000000 & 0.000000 & 1.000000 \\
100000 & 20 & 1.000000 & 0.000000 & 1.000000 & 0.000000 & 1.000000 & 0.000000 & 0.000000 & 1.000000 \\
100000 & 25 & 1.000000 & 0.000000 & 1.000000 & 0.000000 & 1.000000 & 0.000000 & 0.000000 & 1.000000 \\
\bottomrule
\end{tabular}

\vspace{2mm}
\caption{Comparison of VS-$k$NN and Lasso accuracy (ACC) and false positive rate (FPR) for feature selection targeting the Mean and Variance in scenario 4 with $\alpha$ = 0.01.}
\label{tab:scenario4alpha001}
\end{table}

\begin{table}
\centering
\footnotesize
\begin{tabular}{rrrrrrrrrr}
\toprule
N & P & $\text{ACC}^{\mu}_{k\text{-NN}}$ & $\text{FPR}^{\mu}_{k\text{-NN}}$ & $\text{ACC}^{\mu}_{\text{Lasso}}$ & $\text{FPR}^{\mu}_{\text{Lasso}}$ & $\text{ACC}^{\sigma^2}_{k\text{-NN}}$ & $\text{FPR}^{\sigma^2}_{k\text{-NN}}$ & $\text{ACC}^{\sigma^2}_{\text{Lasso}}$ & $\text{FPR}^{\sigma^2}_{\text{Lasso}}$ \\
\midrule
5000 & 5 & 1.000000 & 0.000000 & 1.000000 & 0.000000 & 0.935000 & 0.204000 & 1.000000 & 0.000000 \\
5000 & 10 & 1.000000 & 0.000000 & 1.000000 & 0.000000 & 0.905000 & 0.548000 & 1.000000 & 0.000000 \\
5000 & 20 & 1.000000 & 0.000000 & 1.000000 & 0.000000 & 0.935000 & 0.756000 & 1.000000 & 0.000000 \\
5000 & 25 & 1.000000 & 0.000000 & 1.000000 & 0.000000 & 0.935000 & 0.792800 & 1.000000 & 0.000000 \\
10000 & 5 & 1.000000 & 0.000000 & 1.000000 & 0.000000 & 0.820000 & 0.152000 & 1.000000 & 0.000000 \\
10000 & 10 & 1.000000 & 0.000000 & 1.000000 & 0.000000 & 0.835000 & 0.540000 & 1.000000 & 0.000000 \\
10000 & 20 & 1.000000 & 0.000000 & 1.000000 & 0.000000 & 0.760000 & 0.670000 & 1.000000 & 0.000000 \\
10000 & 25 & 1.000000 & 0.000000 & 1.000000 & 0.000000 & 0.830000 & 0.775400 & 1.000000 & 0.000000 \\
20000 & 5 & 1.000000 & 0.000000 & 1.000000 & 0.000000 & 0.750000 & 0.124000 & 1.000000 & 0.000000 \\
20000 & 10 & 1.000000 & 0.000000 & 1.000000 & 0.000000 & 0.635000 & 0.352000 & 1.000000 & 0.000000 \\
20000 & 20 & 1.000000 & 0.000000 & 1.000000 & 0.000000 & 0.660000 & 0.492000 & 1.000000 & 0.000000 \\
20000 & 25 & 1.000000 & 0.000000 & 1.000000 & 0.000000 & 0.705000 & 0.694200 & 1.000000 & 0.000000 \\
50000 & 5 & 1.000000 & 0.000000 & 1.000000 & 0.000000 & 0.540000 & 0.028000 & 1.000000 & 0.000000 \\
50000 & 10 & 1.000000 & 0.000000 & 1.000000 & 0.000000 & 0.535000 & 0.151800 & 1.000000 & 0.000000 \\
50000 & 20 & 1.000000 & 0.000000 & 1.000000 & 0.000000 & 0.530000 & 0.249200 & 1.000000 & 0.000000 \\
50000 & 25 & 1.000000 & 0.000000 & 1.000000 & 0.000000 & 0.500000 & 0.213800 & 1.000000 & 0.000000 \\
100000 & 5 & 1.000000 & 0.000000 & 1.000000 & 0.000000 & 0.720000 & 0.000000 & 1.000000 & 0.000000 \\
100000 & 10 & 1.000000 & 0.000000 & 1.000000 & 0.000000 & 0.745000 & 0.013000 & 1.000000 & 0.000000 \\
100000 & 20 & 1.000000 & 0.000000 & 1.000000 & 0.000000 & 0.765000 & 0.077000 & 1.000000 & 0.000000 \\
100000 & 25 & 1.000000 & 0.000000 & 1.000000 & 0.000000 & 0.770000 & 0.107200 & 1.000000 & 0.000000 \\
\bottomrule
\end{tabular}

\vspace{2mm}
\caption{Comparison of VS-$k$NN and Lasso accuracy (ACC) and false positive rate (FPR) for feature selection targeting the Mean and Variance in scenario 5 with $\alpha$ = 0.01.}
\label{tab:scenario5alpha001}
\end{table}

\begin{table}
\centering
\footnotesize
\begin{tabular}{rrrrrrrrrr}
\toprule
N & P & $\text{ACC}^{\mu}_{k\text{-NN}}$ & $\text{FPR}^{\mu}_{k\text{-NN}}$ & $\text{ACC}^{\mu}_{\text{Lasso}}$ & $\text{FPR}^{\mu}_{\text{Lasso}}$ & $\text{ACC}^{\sigma^2}_{k\text{-NN}}$ & $\text{FPR}^{\sigma^2}_{k\text{-NN}}$ & $\text{ACC}^{\sigma^2}_{\text{Lasso}}$ & $\text{FPR}^{\sigma^2}_{\text{Lasso}}$ \\
\midrule
5000 & 5 & 1.000000 & 0.000000 & 1.000000 & 0.000000 & 0.900000 & 0.478000 & 0.000000 & 1.000000 \\
5000 & 10 & 1.000000 & 0.000000 & 1.000000 & 0.000000 & 0.870000 & 0.648000 & 0.000000 & 1.000000 \\
5000 & 20 & 1.000000 & 0.000000 & 1.000000 & 0.000000 & 0.900000 & 0.720000 & 0.000000 & 1.000000 \\
5000 & 25 & 0.986800 & 0.000000 & 1.000000 & 0.000000 & 0.850000 & 0.740800 & 0.000000 & 1.000000 \\
10000 & 5 & 1.000000 & 0.000000 & 1.000000 & 0.000000 & 0.710000 & 0.178000 & 0.000000 & 1.000000 \\
10000 & 10 & 1.000000 & 0.000000 & 1.000000 & 0.000000 & 0.790000 & 0.362600 & 0.000000 & 1.000000 \\
10000 & 20 & 1.000000 & 0.000000 & 1.000000 & 0.000000 & 0.730000 & 0.402000 & 0.000000 & 1.000000 \\
10000 & 25 & 1.000000 & 0.000000 & 1.000000 & 0.000000 & 0.770000 & 0.409800 & 0.000000 & 1.000000 \\
20000 & 5 & 1.000000 & 0.000000 & 1.000000 & 0.000000 & 0.840000 & 0.042600 & 0.000000 & 1.000000 \\
20000 & 10 & 1.000000 & 0.000000 & 1.000000 & 0.000000 & 0.820000 & 0.016000 & 0.000000 & 1.000000 \\
20000 & 20 & 1.000000 & 0.000000 & 1.000000 & 0.000000 & 0.890000 & 0.095200 & 0.000000 & 1.000000 \\
20000 & 25 & 1.000000 & 0.000000 & 1.000000 & 0.000000 & 0.830000 & 0.041600 & 0.000000 & 1.000000 \\
50000 & 5 & 1.000000 & 0.000000 & 1.000000 & 0.000000 & 1.000000 & 0.006600 & 0.000000 & 1.000000 \\
50000 & 10 & 1.000000 & 0.000000 & 1.000000 & 0.000000 & 1.000000 & 0.000000 & 0.000000 & 1.000000 \\
50000 & 20 & 1.000000 & 0.000000 & 1.000000 & 0.000000 & 0.990000 & 0.013200 & 0.000000 & 1.000000 \\
50000 & 25 & 1.000000 & 0.000000 & 1.000000 & 0.000000 & 1.000000 & 0.056200 & 0.000000 & 1.000000 \\
100000 & 5 & 1.000000 & 0.000000 & 1.000000 & 0.000000 & 1.000000 & 0.000000 & 0.000000 & 1.000000 \\
100000 & 10 & 1.000000 & 0.000000 & 1.000000 & 0.000000 & 1.000000 & 0.000000 & 0.000000 & 1.000000 \\
100000 & 20 & 1.000000 & 0.000000 & 1.000000 & 0.000000 & 1.000000 & 0.006600 & 0.000000 & 1.000000 \\
100000 & 25 & 1.000000 & 0.000000 & 1.000000 & 0.000000 & 1.000000 & 0.026400 & 0.000000 & 1.000000 \\
\bottomrule
\end{tabular}

\vspace{2mm}
\caption{Comparison of VS-$k$NN and Lasso accuracy (ACC) and false positive rate (FPR) for feature selection targeting the Mean and Variance in scenario 6 with $\alpha$ = 0.01.}
\label{tab:scenario6alpha001}
\end{table}

\begin{table}
\centering
\footnotesize
\begin{tabular}{rrrrrrrrrr}
\toprule
N & P & $\text{ACC}^{\mu}_{k\text{-NN}}$ & $\text{FPR}^{\mu}_{k\text{-NN}}$ & $\text{ACC}^{\mu}_{\text{Lasso}}$ & $\text{FPR}^{\mu}_{\text{Lasso}}$ & $\text{ACC}^{\sigma^2}_{k\text{-NN}}$ & $\text{FPR}^{\sigma^2}_{k\text{-NN}}$ & $\text{ACC}^{\sigma^2}_{\text{Lasso}}$ & $\text{FPR}^{\sigma^2}_{\text{Lasso}}$ \\
\midrule
5000 & 5 & 0.835000 & 0.004000 & 1.000000 & 0.000000 & 0.795000 & 0.160000 & 0.750000 & 0.250000 \\
5000 & 10 & 0.765000 & 0.000000 & 1.000000 & 0.000000 & 0.790000 & 0.522000 & 0.750000 & 0.250000 \\
5000 & 20 & 0.780000 & 0.000000 & 1.000000 & 0.000000 & 0.835000 & 0.644000 & 0.750000 & 0.250000 \\
5000 & 25 & 0.735000 & 0.067200 & 1.000000 & 0.000000 & 0.820000 & 0.735200 & 0.750000 & 0.250000 \\
10000 & 5 & 0.995000 & 0.000000 & 1.000000 & 0.000000 & 0.855000 & 0.160000 & 0.750000 & 0.250000 \\
10000 & 10 & 0.995000 & 0.000000 & 1.000000 & 0.000000 & 0.800000 & 0.536000 & 0.750000 & 0.250000 \\
10000 & 20 & 0.995000 & 0.000000 & 1.000000 & 0.000000 & 0.795000 & 0.722000 & 0.750000 & 0.250000 \\
10000 & 25 & 0.990000 & 0.000000 & 1.000000 & 0.000000 & 0.775000 & 0.688200 & 0.750000 & 0.250000 \\
20000 & 5 & 1.000000 & 0.000000 & 1.000000 & 0.000000 & 0.690000 & 0.136000 & 0.750000 & 0.250000 \\
20000 & 10 & 1.000000 & 0.000000 & 1.000000 & 0.000000 & 0.715000 & 0.380000 & 0.750000 & 0.250000 \\
20000 & 20 & 1.000000 & 0.000000 & 1.000000 & 0.000000 & 0.715000 & 0.556600 & 0.750000 & 0.250000 \\
20000 & 25 & 1.000000 & 0.000000 & 1.000000 & 0.000000 & 0.690000 & 0.635600 & 0.750000 & 0.250000 \\
50000 & 5 & 1.000000 & 0.000000 & 1.000000 & 0.000000 & 0.545000 & 0.029000 & 0.750000 & 0.250000 \\
50000 & 10 & 1.000000 & 0.000000 & 1.000000 & 0.000000 & 0.415000 & 0.110000 & 0.750000 & 0.250000 \\
50000 & 20 & 1.000000 & 0.000000 & 1.000000 & 0.000000 & 0.555000 & 0.238200 & 0.750000 & 0.250000 \\
50000 & 25 & 1.000000 & 0.000000 & 1.000000 & 0.000000 & 0.530000 & 0.175400 & 0.750000 & 0.250000 \\
100000 & 5 & 1.000000 & 0.000000 & 1.000000 & 0.000000 & 0.765000 & 0.008000 & 0.750000 & 0.250000 \\
100000 & 10 & 1.000000 & 0.000000 & 1.000000 & 0.000000 & 0.745000 & 0.021000 & 0.750000 & 0.250000 \\
100000 & 20 & 1.000000 & 0.000000 & 1.000000 & 0.000000 & 0.780000 & 0.079200 & 0.750000 & 0.250000 \\
100000 & 25 & 1.000000 & 0.000000 & 1.000000 & 0.000000 & 0.755000 & 0.073000 & 0.750000 & 0.250000 \\
\bottomrule
\end{tabular}

\vspace{2mm}
\caption{Comparison of VS-$k$NN and Lasso accuracy (ACC) and false positive rate (FPR) for feature selection targeting the Mean and Variance in scenario 7 with $\alpha$ = 0.01.}
\label{tab:scenario7alpha001}
\end{table}

\begin{table}
\centering
\footnotesize
\begin{tabular}{rrrrrrrrrr}
\toprule
N & P & $\text{ACC}^{\mu}_{k\text{-NN}}$ & $\text{FPR}^{\mu}_{k\text{-NN}}$ & $\text{ACC}^{\mu}_{\text{Lasso}}$ & $\text{FPR}^{\mu}_{\text{Lasso}}$ & $\text{ACC}^{\sigma^2}_{k\text{-NN}}$ & $\text{FPR}^{\sigma^2}_{k\text{-NN}}$ & $\text{ACC}^{\sigma^2}_{\text{Lasso}}$ & $\text{FPR}^{\sigma^2}_{\text{Lasso}}$ \\
\midrule
5000 & 10 & 1.000000 & 0.000000 & 1.000000 & 0.000000 & 1.000000 & 0.000000 & 0.000000 & 1.000000 \\
5000 & 25 & 0.871600 & 0.000000 & 1.000000 & 0.000000 & 1.000000 & 0.000000 & 0.000000 & 1.000000 \\
5000 & 50 & 0.611000 & 0.000000 & 1.000000 & 0.000000 & 1.000000 & 0.000000 & 0.000000 & 1.000000 \\
5000 & 100 & 0.499800 & 0.073600 & 1.000000 & 0.000000 & 1.000000 & 0.000000 & 0.000000 & 1.000000 \\
10000 & 10 & 1.000000 & 0.000000 & 1.000000 & 0.000000 & 1.000000 & 0.000000 & 0.000000 & 1.000000 \\
10000 & 25 & 1.000000 & 0.000000 & 1.000000 & 0.000000 & 1.000000 & 0.000000 & 0.000000 & 1.000000 \\
10000 & 50 & 1.000000 & 0.000000 & 1.000000 & 0.000000 & 1.000000 & 0.000000 & 0.000000 & 1.000000 \\
10000 & 100 & 0.943200 & 0.000000 & 1.000000 & 0.000000 & 1.000000 & 0.000000 & 0.000000 & 1.000000 \\
20000 & 10 & 1.000000 & 0.000000 & 1.000000 & 0.000000 & 1.000000 & 0.000000 & 0.000000 & 1.000000 \\
20000 & 25 & 1.000000 & 0.000000 & 1.000000 & 0.000000 & 1.000000 & 0.000000 & 0.000000 & 1.000000 \\
20000 & 50 & 1.000000 & 0.000000 & 1.000000 & 0.000000 & 1.000000 & 0.000000 & 0.000000 & 1.000000 \\
20000 & 100 & 1.000000 & 0.000000 & 1.000000 & 0.000000 & 1.000000 & 0.000000 & 0.000000 & 1.000000 \\
50000 & 10 & 1.000000 & 0.000000 & 1.000000 & 0.000000 & 1.000000 & 0.000000 & 0.000000 & 1.000000 \\
50000 & 25 & 1.000000 & 0.000000 & 1.000000 & 0.000000 & 1.000000 & 0.000000 & 0.000000 & 1.000000 \\
50000 & 50 & 1.000000 & 0.000000 & 1.000000 & 0.000000 & 1.000000 & 0.000000 & 0.000000 & 1.000000 \\
50000 & 100 & 1.000000 & 0.000000 & 1.000000 & 0.000000 & 1.000000 & 0.000000 & 0.000000 & 1.000000 \\
100000 & 10 & 1.000000 & 0.000000 & 1.000000 & 0.000000 & 1.000000 & 0.000000 & 0.000000 & 1.000000 \\
100000 & 25 & 1.000000 & 0.000000 & 1.000000 & 0.000000 & 1.000000 & 0.000000 & 0.000000 & 1.000000 \\
100000 & 50 & 1.000000 & 0.000000 & 1.000000 & 0.000000 & 1.000000 & 0.000000 & 0.000000 & 1.000000 \\
100000 & 100 & 1.000000 & 0.000000 & 1.000000 & 0.000000 & 1.000000 & 0.000000 & 0.000000 & 1.000000 \\
\bottomrule
\end{tabular}

\vspace{2mm}
\caption{Comparison of VS-$k$NN and Lasso accuracy (ACC) and false positive rate (FPR) for feature selection targeting the Mean and Variance in scenario 8 with $\alpha$ = 0.01.}
\label{tab:scenario8alpha001}
\end{table}

\begin{table}
\centering
\footnotesize
\begin{tabular}{rrrrrrrrrr}
\toprule
N & P & $\text{ACC}^{\mu}_{k\text{-NN}}$ & $\text{FPR}^{\mu}_{k\text{-NN}}$ & $\text{ACC}^{\mu}_{\text{Lasso}}$ & $\text{FPR}^{\mu}_{\text{Lasso}}$ & $\text{ACC}^{\sigma^2}_{k\text{-NN}}$ & $\text{FPR}^{\sigma^2}_{k\text{-NN}}$ & $\text{ACC}^{\sigma^2}_{\text{Lasso}}$ & $\text{FPR}^{\sigma^2}_{\text{Lasso}}$ \\
\midrule
5000 & 10 & 0.000000 & 1.000000 & 1.000000 & 0.000000 & 0.894400 & 0.196000 & 0.000000 & 0.000000 \\
5000 & 25 & 0.000000 & 1.000000 & 1.000000 & 0.000000 & 0.929800 & 0.645600 & 0.000000 & 0.000000 \\
5000 & 50 & 0.000000 & 1.000000 & 1.000000 & 0.000000 & 0.902400 & 0.836000 & 0.000000 & 0.000000 \\
5000 & 100 & 0.000000 & 1.000000 & 1.000000 & 0.000000 & 0.904800 & 0.888000 & 0.000000 & 0.000000 \\
10000 & 10 & 0.000000 & 1.000000 & 1.000000 & 0.000000 & 0.773800 & 0.148000 & 0.000000 & 0.000000 \\
10000 & 25 & 0.000000 & 1.000000 & 1.000000 & 0.000000 & 0.759400 & 0.639800 & 0.000000 & 0.000000 \\
10000 & 50 & 0.000000 & 1.000000 & 1.000000 & 0.000000 & 0.734600 & 0.791400 & 0.000000 & 0.000000 \\
10000 & 100 & 0.000000 & 1.000000 & 1.000000 & 0.000000 & 0.884800 & 0.899600 & 0.000000 & 0.000000 \\
20000 & 10 & 0.000000 & 1.000000 & 1.000000 & 0.000000 & 0.829200 & 0.160000 & 0.000000 & 0.000000 \\
20000 & 25 & 0.000000 & 1.000000 & 1.000000 & 0.000000 & 0.658600 & 0.545000 & 0.000000 & 0.000000 \\
20000 & 50 & 0.000000 & 1.000000 & 1.000000 & 0.000000 & 0.584000 & 0.768800 & 0.000000 & 0.000000 \\
20000 & 100 & 0.000000 & 1.000000 & 1.000000 & 0.000000 & 0.677000 & 0.900800 & 0.000000 & 0.000000 \\
50000 & 10 & 0.000000 & 1.000000 & 1.000000 & 0.000000 & 0.621200 & 0.148000 & 0.000000 & 0.000000 \\
50000 & 25 & 0.000000 & 1.000000 & 1.000000 & 0.000000 & 0.403800 & 0.584000 & 0.000000 & 0.000000 \\
50000 & 50 & 0.000000 & 1.000000 & 1.000000 & 0.000000 & 0.348400 & 0.717600 & 0.000000 & 0.000000 \\
50000 & 100 & 0.000000 & 1.000000 & 1.000000 & 0.000000 & 0.476000 & 0.819200 & 0.000000 & 0.000000 \\
100000 & 10 & 0.000000 & 1.000000 & 1.000000 & 0.000000 & 0.525800 & 0.094600 & 0.000000 & 0.000000 \\
100000 & 25 & 0.000000 & 1.000000 & 1.000000 & 0.000000 & 0.427000 & 0.448200 & 0.000000 & 0.000000 \\
100000 & 50 & 0.000000 & 1.000000 & 1.000000 & 0.000000 & 0.265200 & 0.518600 & 0.000000 & 0.000000 \\
100000 & 100 & 0.000000 & 1.000000 & 1.000000 & 0.000000 & 0.168000 & 0.702200 & 0.000000 & 0.000000 \\
\bottomrule
\end{tabular}

\vspace{2mm}
\caption{Comparison of VS-$k$NN and Lasso accuracy (ACC) and false positive rate (FPR) for feature selection targeting the Mean and Variance in scenario 9 with $\alpha$ = 0.01.}
\label{tab:scenario9alpha001}
\end{table}

\end{document}